\documentclass[11pt,onecolumn,english,showpacs,preprintnumbers,amsmath,amssymb,floatfix,nofootinbib,superscriptaddress]{revtex4-1}

\usepackage{epsfig}
\usepackage{graphics}
\usepackage{latexsym}
\usepackage{amsmath}
\usepackage{amssymb}
\usepackage{rotating}
\usepackage{subfigure}
\usepackage{bm}
\usepackage{color}
\usepackage{ulem}
\usepackage{booktabs}
\usepackage[colorlinks = true,
linkcolor = magenta,
urlcolor  = blue,
citecolor = red,
anchorcolor = blue]{hyperref}
\usepackage{extsizes}
\usepackage{geometry}
\usepackage{pxfonts}
\usepackage{graphicx}
\usepackage{booktabs}
\usepackage[table]{xcolor}
\definecolor{lightgray}{gray}{0.85}
\usepackage{color}

\begin{document}

\preprint{DESY~19-066\hspace{13.cm} ISSN~0418-9833}
\preprint{April 2019\hspace{15.9cm}}

%
%%%%%%%%%%%%%%%%%%%%%%%%%%%%%%%%%%%%%%%%%%%%%%%%%%%%%%%%%%%%%%%%%%%%%%%%%%%%%%%%%%%%%%%%%%%%%%%%%%%%%%%%%%%%%%%%%%%%%%%%%%%%%%%%%%%%%%%
%
\title{ $B$-hadron fragmentation functions at next-to-next-to-leading order from global analysis of $e^+e^-$ annihilation data }
%
%%%%%%%%%%%%%%%%%%%%%%%%%%%%%%%%%%%%%%%%%%%%%%%%%%%%%%%%%%%%%%%%%%%%%%%%%%%%%%%%%%%%%%%%%%%%%%%%%%%%%%%%%%%%%%%%%%%%%%%%%%%%%%%%%%%%%%%
%

\author{Maral Salajegheh}
\email{M.Salajegheh@stu.yazd.ac.ir}
\affiliation{Faculty of Physics, Yazd University, P.~O.~Box~89195-741, Yazd,
  Iran}
% University Boulevard, Safayieh
% https://www.yazd.ac.ir
\author{S. Mohammad Moosavi Nejad}
\email{Mmoosavi@yazd.ac.ir}
\affiliation{Faculty of Physics, Yazd University, P.~O.~Box~89195-741, Yazd,
  Iran}
\affiliation{School of Particles and Accelerators, Institute for Research in
  Fundamental Sciences (IPM), P.~O.~Box 19395-5531, Tehran, Iran}
% Niavaran Square
% http://www.ipm.ac.ir/
\author{Hamzeh Khanpour}
\email{Hamzeh.Khanpour@mail.ipm.ir}
\affiliation{Department of Physics, University of Science and Technology of
	Mazandaran, P.~O.~Box 48518-78195, Behshahr, Iran}
\affiliation{School of Particles and Accelerators, Institute for Research in
	Fundamental Sciences (IPM), P.~O.~Box 19395-5531, Tehran, Iran}
% Pasdaran Street 47415
%http://www.mazust.ac.ir
\author{Bernd A. Kniehl}
\email{kniehl@desy.de}
\affiliation{{II.}~Institut f\"ur Theoretische Physik, Universit\"at Hamburg,
  Luruper Chaussee 149, 22761 Hamburg, Germany}
\author{Maryam Soleymaninia}
\email{Maryam\_Soleymaninia@ipm.ir}
\affiliation{Department of Physics, Shahid Rajaee Teacher Training University,
	Lavizan, Tehran 16788, Iran}
\affiliation{School of Particles and Accelerators, Institute for Research in
	Fundamental Sciences (IPM), P.~O.~Box 19395-5531, Tehran, Iran}
% P.O. Box 16785-163, Lavizan, Tehran, IRAN
% http://www.srttu.edu/

\date{\today}

%
%%%%%%%%%%%%%%%%%%%%%%%%%%%%%%%%%%%%%%%%%%%%%%%%%%%%%%%%%%%%%%%%%%%%%%%%%
\begin{abstract}
  We present nonperturbative fragmentation functions (FFs) for bottom-flavored
  ($B$) hadrons both at next-to-leading (NLO) and, for the first time, at
  next-to-next-to-leading order (NNLO) in the $\overline{\mathrm{MS}}$
  factorization scheme with five massless quark flavors.
  They are determined by fitting all available experimental data of inclusive
  single $B$-hadron production in $e^+e^-$ annihilation, from the ALEPH, DELPHI,
  and OPAL Collaborations at CERN LEP1 and the SLD Collaboration at SLAC SLC.
  The uncertainties in these FFs as well as in the corresponding observables
  are estimated using the Hessian approach.
  We perform comparisons with available NLO sets of $B$-hadron FFs.
  We apply our new FFs to generate theoretical predictions for the energy
  distribution of $B$ hadrons produced through the decay of unpolarized or
  polarized top quarks, to be measured at the CERN LHC. 
\end{abstract}
%

%13.38.Dg 	Decays of Z bosons
%13.66.Bc 	Hadron production in e−e+ interactions
%13.85.Ni 	Inclusive production with identified hadrons
%13.87.Fh 	Fragmentation into hadrons 
%14.40.Nd 	Bottom mesons
%14.65.Ha 	Top quark   
             
\maketitle
\tableofcontents{}

%
%%%%%%%%%%%%%%%%%%%%%%%%%%%%%%%%%%%%%%%%%%%%%%%%%%%%%%%%%%%%%%%%%%%%%%%
\section{Introduction}\label{sec:introduction}
%%%%%%%%%%%%%%%%%%%%%%%%%%%%%%%%%%%%%%%%%%%%%%%%%%%%%%%%%%%%%%%%%%%%%%%
%

For a long time, there has been considerable interest in the study of
bottom-flavored-hadron ($B$-hadron) production at hadron and $e^+e^-$ colliders,
both experimentally and theoretically.
Historically, the first measurements were performed more than three decades
ago by the UA1 Collaboration at the CERN $Sp\bar{p}S$ collider
\cite{Albajar:1988th} operating at center-of-mass (CM) energy
$\sqrt{s} = 630$~GeV.

In the framework of the parton model of QCD, the description of the inclusive
single production of identified hadrons $h$ involves fragmentation functions
(FFs), $D_a^h(x,Q^2)$.
At leading order, their values correspond to the probability that the
colored parton $a$, which is produced at short distance, of order
$1/\sqrt{Q^2}$, fragments into the colorless hadron $h$ carrying the fraction
$x$ of the energy of $a$.
Given their $x$ dependence at some scale $Q_0$, the evolution of the FFs with
$Q^2$ may be computed perturbatively from the timelike
Dokshitzer-Gribov-Lipatov-Altarelli-Parisi (DGLAP)
equations \cite{Gribov:1972ri,Altarelli:1977zs,Dokshitzer:1977sg}.
The timelike splitting functions $P_{a\to b}^T(x,\alpha_s(Q^2))$ appearing
therein are known through NNLO
\cite{Mitov:2006ic,Moch:2007tx,Almasy:2011eq}.
In the case of $e^+e^-$ annihilation, the hard-scattering cross sections for
the inclusive production of parton $a$, to be convoluted with $D_a^h(x,Q^2)$,
are also known through NNLO
\cite{Mitov:2006ic,Rijken:1996vr,Rijken:1996npa,Rijken:1996ns,Mitov:2006wy}.
This allows one to interpret $e^+e^-$ data of the inclusive single production
of hadron $h$ at NNLO and thus to extract FFs at this order
\cite{Anderle:2015lqa,Bertone:2017tyb,Soleymaninia:2018uiv,%
Soleymaninia:2019sjo}.
Owing to the factorization theorem, the FFs are independent of the process by
which parton $a$ is produced.
This allows one to transfer experimental information from $e^+e^-$
annihilation to any other production mechanism, such as photoproduction,
leptoproduction, hadroproduction, and two-photon scattering.
Of all these processes, $e^+e^-$ annihilation provides the cleanest laboratory
for the extraction of FFs, being devoid of nonperturbative effects beyond
fragmentation itself.
Presently, there is particular interest in hadroproduction at the BNL
Relativistic Heavy Ion Collider (RHIC) and the CERN Large Hadron Collider
(LHC) due to ongoing experiments.

The parton model of QCD implemented in the modified minimal-subtraction
($\overline{\mathrm{MS}}$) factorization scheme with $n_f$ massless-quark
flavors, the so-called zero-mass variable-flavor-number scheme (ZM-VFNS), can
also be applied to the open production of heavy flavors, such as $D$ and $B$
hadrons, provided the hard energy scale characteristic for the production
process is sufficiently larger than the heavy-flavor mass.
This is certainly the case for all the applications here, because
$M_B\ll M_Z, m_t$.
Recently, $D$ hadron FFs have been provided at NNLO in
Ref.~\cite{Soleymaninia:2017xhc}.
Here, we perform the first NNLO determination of $B$-hadron FFs.

In Refs.~\cite{Binnewies:1998vm,Kniehl:2008zza}, $B$-hadron FFs were extracted
at NLO in the ZM-VFNS by fitting to the fractional-energy distributions
$d\sigma/dx_B$ of the cross section of $e^+e^-\rightarrow B+X$ measured by
the ALEPH~\cite{Heister:2001jg} and OPAL~\cite{Abbiendi:2002vt} Collaborations
at the CERN Large Electron Positron Collider (LEP1) and the SLD
Collaboration~\cite{Abe:2002iq} at the SLAC Linear Collider (SLC).

In the meantime, also the DELPHI Collaboration have reported a similar
measurement at LEP1 \cite{DELPHI:2011aa}.
In the present work, these data are, for the first time, included in a
$B$-hadron FF fit.
On the other hand, we are not aware of any other such data from $e^+e^-$
annihilation.
In want of NNLO hard-scattering cross sections for inclusive single $B$-hadron
production from other initial states, we do concentrate here on $e^+e^-$
annihilation.
We also go beyond Refs.~\cite{Binnewies:1998vm,Kniehl:2008zza} by performing
a full-fledged error estimation, both for the FFs and the resulting
differential cross sections, using the Hessian approach \cite{Pumplin:2000vx}.

The LEP1 experiments \cite{Heister:2001jg,Abbiendi:2002vt,DELPHI:2011aa}
identified the $B$ hadrons by their semileptonic decays,
$B\to D^{(\ast)}\ell\nu$, while the SLD Collaboration \cite{Abe:2002iq}
collected an inclusive sample of reconstructed $B$-hadron decay vertices.
The bulk of the experimentally observed $B$ hadrons is made up by
$B^\pm$, $B^0/\overline{B}^{\,0}$, and $B_s^0/\overline{B}_s^{\,0}$ mesons.

The outline of this paper is as follows:
In Section~\ref{sec:framework}, we describe the theoretical framework of
inclusive single hadron production in $e^+e^-$ annihilation through NNLO in
the ZM-FVNS and introduce our parametrization of the  $b/\bar{b}\to B$ FF at
the initial scale.
In Section~\ref{sec:errorcalculation}, we explain the minimization method in
our analysis and our approach to error estimation.
In Section.~\ref{sec:results}, our NLO and NNLO results are presented and
compared with the experimental data fitted to.
In Section.~\ref{sec:B-meson-LHC}, we present our NLO predictions for the
normalized-energy distributions of $B$ hadrons from decays of (un)polarized
top quarks. Our conclusions are given in Section~\ref{sec:conclusion}.

%%%%%%%%%%%%%%%%%%%%%%%%%%%%%%%%%%%%%%%%%%%%%%%%%%%%%%%%%%%%%%%%%%%%%%%
\section{QCD framework for $B$-hadron FFs} \label{sec:framework}
%%%%%%%%%%%%%%%%%%%%%%%%%%%%%%%%%%%%%%%%%%%%%%%%%%%%%%%%%%%%%%%%%%%%%%%

As mentioned in Sec.~\ref{sec:introduction}, we fit nonperturbative $B$-hadron
FFs to measured $x_B$ distributions of the cross section of
%---------------------------------------
\begin{eqnarray}\label{one}
e^+ e^- \to (\gamma^\star, Z) \to B + X,	
\end{eqnarray}
%---------------------------------------
where $X$ refers to the unobserved part of the final state.
In the following, we explain how to evaluate the cross section of process
(\ref{one}) at NLO and NNLO in the ZM-VFNS.
Denoting the four-momenta of the virtual gauge boson and the $B$ hadron by $q$
and $p_B$, respectively, we have $s = q^2$, $p_B^2 = m_B^2$, and
$x_B = 2(p_B \cdot q)/q^2$.
In the CM frame, $x_B = 2E_B/\sqrt{s}$ is the energy of the $B$
hadron in units of the beam energy.
In the ZM-VFNS, we have
%------------------------------------------------------
\begin{equation}\label{convolution}
\frac{1}{\sigma_{\rm tot}}\,\frac{d\sigma}{dx_B}(e^+e^-\rightarrow B+X)=
\sum_i \int_{x_B}^1
\frac{dx_i}{x_i}\,\frac{1}{\sigma_{\rm tot}}\,
\frac{d\sigma_i}{dx_i}(x_i, \mu_R, \mu_F)D_i^B(\frac{x_B}{x_i}, \mu_F),
\end{equation}
%------------------------------------------------------
where $i=g, u, \bar{u}, \ldots, b, \bar{b}$ runs over the active partons
with four-momenta $p_i$,
$d\sigma_i(x_i, \mu_R, \mu_F)/dx_i$ is the partonic cross section of
$e^+e^- \to i+X$ differential in $x_i= 2(p_i \cdot q)/q^2$,
$D_i^B(z, \mu_F)$ is the $i\to B$ FF, and $\mu_R$ and $\mu_F$ are the
renormalization and factorization scales, respectively.
The latter are a priori arbitrary, but a typical choice is
$\mu_F = \mu_R = \sqrt{s}$.
In the CM frame, $z = x_B/x_i$ is the fraction of energy passed on from parton
$i$ to the $B$ hadron.
It is customary in experimental analyses to normalize Eq.~\eqref{convolution}
by the total hadronic cross section,
%------------------------------------------------------
\begin{eqnarray}\label{sigmatotal}
\sigma_{\rm tot} = 
\frac{4\pi\alpha^2(s)}{s}
\left(\sum_i^{n_f}\tilde{e}_i^2(s)\right)
\left(1+\alpha_s K_{\rm QCD}^{(1)}+
\alpha_s^2K_{\rm QCD}^{(2)} +
\cdots\right),
\end{eqnarray}
%------------------------------------------------------
where $\alpha$ and $\alpha_s$ are the fine-structure and strong-coupling
constants, respectively, $\tilde{e}_i$ is the effective electroweak charge of
quark $i$, and the coefficient $K_{\rm QCD}^{(n)}$ contains the N${}^n$LO
correction.
Here, we need $K_{\rm QCD}^{(1)}=3C_F/(4\pi)$ and $K_{\rm QCD}^{(2)}\approx1.411$
\cite{Chetyrkin:1979bj}.

The $z$ distribution of the $b \to B$ FF at the starting scale $\mu_0$ is a
genuinely nonperturbative quantity to be extracted from experimental data.
Its form is unknown, and an educated guess is in order.
The selection criterion is to score a minimum $\chi^2$ value as small as
possible with a set of fit parameters as minimal as possible.
As in Refs.~\cite{Binnewies:1998vm,Kniehl:2008zza}, we adopt here the
simple power ansatz \cite{Kartvelishvili:1985ac},
%------------------------------------------------------
\begin{equation}\label{input}
D_b^B(z, \mu_0)= N_bz^{\alpha_b}(1 - z)^{\beta_b},
\end{equation}
%------------------------------------------------------
with fit parameters $N_b$, $\alpha_b$, and $\beta_b$, and choose
$\mu _0 = m_b = 4.5$~GeV.
This ansatz was found to enable excellent fits~\cite{Binnewies:1998vm,Kniehl:2008zza}.
The $i\to B$ FFs for the other quarks and the gluon are assumed to be zero at
$\mu_F=\mu_0$ and are generated through the DGLAP evolution to larger values of
$\mu_F$.
We take $\alpha _s(M_Z)$ to be an input parameter and adopt the world average
value 0.1181 for $n_f=5$ \cite{Tanabashi:2018oca} both at NLO and NNLO.

As mentioned in Sec.~\ref{sec:introduction}, we fit to ALEPH
\cite{Heister:2001jg}, DELPHI \cite{DELPHI:2011aa}, OPAL
\cite{Abbiendi:2002vt}, and SLD \cite{Abe:2002iq} data.
These data sets reach down to very small $x_B$ values, which fall outside the
range of validity of our fixed-order approach.
In fact, in the small-$x_B$ limit, both the timelike splitting functions and
the hard-scattering cross sections develop soft-gluon logarithms that require
resummation.
At the same time, finite-$m_b$ and finite-$M_B$ effects become relevant there.
We leave the implementation of these refinements for future work, and instead
impose appropriate minimum-$x_B$ cuts for the time being.
Specifically, we only include ALEPH data points with $x_B \geq 0.25$,
DELPHI data points with $x_B \geq 0.36$, OPAL data points with
$x_B \geq 0.325$, and SLD data points with $x_B \geq 0.28$.
This enables acceptable fits within the fit range, at the expense of certain
deviations in the small-$x_B$ range, of course.
The fixed-order approach is also challenged in the large-$x_B$ limit, by the
emergence of threshold logarithms, which also require resummation.
In practice, however, these effects do not jeopardize the quality of our fits,
so that we refrain from imposing maximum-$x_B$ cuts.

%
%%%%%%%%%%%%%%%%%%%%%%%%%%%%%%%%%%%%%%%%%%%%%%%%%%%%%%%%%%%%%%%%%%%%%%%
\section{Determination of $B$-hadron FFs and their uncertainties}\label{sec:errorcalculation}
%%%%%%%%%%%%%%%%%%%%%%%%%%%%%%%%%%%%%%%%%%%%%%%%%%%%%%%%%%%%%%%%%%%%%%%
%

We now explain our fitting procedure.
For a given set $p=\{N_b,\alpha_b,\beta_b\}$ of fit parameters, the goodness of
the overall description of the experimental data by the theoretical predictions
is measured by the global $\chi^2$ value,
%------------------------------------------------------
\begin{equation}\label{eq:chi2}
\chi_{\mathrm{global}}^2(p)
= \sum_{n=1}^{N^{\mathrm{exp}}} w_n\chi_n^2(p) \,,
\end{equation}
%------------------------------------------------------
where $n$ labels the $N^{\mathrm{exp}}=4$ experimental data sets, $w_n$ are their
weight factors \cite{Stump:2001gu,Blumlein:2006be}, which we take to be unity,
and
%------------------------------------------------------
\begin{equation}\label{eq:chi2global}
\chi_n^2 (p) = \left( \frac{1 -{\cal N}_n }
    {\Delta{\cal N}_n}\right)^2 + \sum_{i=1}^{N_n^{\mathrm{data}}}
      \left(\frac{{\cal N}_n {\cal F}_{n,i}^{\mathrm{exp}} - 
{\cal F}_{n,i}^{\mathrm{theo}}(p)}
{{\cal N}_n \Delta {\cal F}_{n,i}^{\mathrm{exp}}}\right)^2
\end{equation}
%------------------------------------------------------
is the $\chi^2$ value of data set $n$.
On the experimental side,
${\cal F}_{n,i}^{\mathrm{exp}}$ is the central value of
$(1/\sigma_{\mathrm{tot}})d\sigma/dx_B$ measured in bin $i$ out of the
$N_n^{\mathrm{data}}$ bins in data set $n$,
$\Delta{\cal F}_{n,i}^{\mathrm{exp}}$ is its individual error obtained by
combining statistical and systematic errors in quadrature, 
${\cal N}_n$ is the unknown overall normalization factor of data set $n$ to
be fitted, and
$\Delta{\cal N}_n$ is its error as quoted by the experimental collaboration.
On the theoretical side, ${\cal F}_{n,i}^{\mathrm{theo}}(p)$ is the respective NLO
or NNLO prediction. 

We determine the fit parameters $p$ by minimizing Eq.~(\ref{eq:chi2}) with the
help of the Monte Carlo package MINUIT \cite{James:1975dr} from the CERN
program library.
We adopt a two-step procedure.
In the pre-fitting stage, we determine the four values ${\cal N}_n$ by fitting
them simultaneously with the three fit parameters $p$.
In the main fitting stage, we then refine the determination of $p$ with large
statistics keeping ${\cal N}_n$ fixed.
In the evaluation of $\chi_{\mathrm{global}}^2(p)/\mathrm{d.o.f.}$, we take the
number of degrees of freedom to be the overall number of data points fitted to
minus three for the proper fit parameters $p$.
We find the APFEL library~\cite{Bertone:2013vaa} to be a very
useful FF fitting tool.

We now describe our methodology for the estimation of the uncertainties in the
$B$-hadron FFs.
We adopt the Hessian approach to the propagation of uncertainties from the
experimental data sets to the FFs, which has proven of value in global
analyses of parton distribution functions and has been frequently applied
there.
For definiteness, we ignore additional sources of uncertainties, which are
mostly of theoretical origin and are negligible against the experimental
uncertainties taken into account here.
In the following, we briefly review our procedure.
For more details, we refer to Ref.~\cite{Martin:2009iq}.

In the Hessian approach, the uncertainty bands on the $B$-hadron FFs,
$\Delta D_b^B(z)$, may be obtained through linear error
propagation,
%--------------------------------
%\begin{equation}\label{eq:uncertainties}
%[\Delta {\cal D}^B (z)]^2 =
%\Delta \chi^2_{\mathrm{global}} 
%\left( \sum_i (\frac{\partial
%\Delta {\cal D}^B
%(z, {\hat \xi})}{\partial
%\xi_i})^2 C_{i i}
%+ \sum_{i \neq j}
%( \frac{\partial \Delta
%{\cal D}^B (z, {\hat \xi})}
%{\partial \xi_i}  C_{i j}  
%\frac{\partial \Delta {\cal D}^B
%(z, {\hat \xi})}{\partial \xi_j} )
%\right ) \,,
%\end{equation}
%------------------------------------------------------
\begin{equation}\label{eq:uncertainties}
[\Delta D^B (z)]^2 =
\Delta \chi^2_{\mathrm{global}}(\hat{p}) 
\sum_{i,j}\frac{\partial\Delta D^B(z,\hat{p})}{\partial p_i}
H_{i j}^{-1}(\hat{p})  
\frac{\partial\Delta D^B(z,\hat{p})}{\partial p_j}\,,
\end{equation}
%------------------------------------------------------
where $p_i$ ($i = 1,2,3$) are the free parameters in Eq.~(\ref{input}),
$\hat{p}_i$ are their optimized values, and
$H^{-1}(p)$ is the covariance matrix, which is a default output of the MINUIT
program \cite{James:1975dr}.
In Eq.~(\ref{eq:uncertainties}), we have suppressed the label $\mu_F$ for the
factorization scale, which we take to be $\mu_0$.
The error bands $\Delta D^B(z)$ are subject to DGLAP evolution in $\mu_F$ along
with the central values $D^B(z)$.
The confidence level (C.L.) is controlled by
$T^2 = \Delta \chi^2_{\mathrm{global}}$.
We adopt the standard parameter fitting criterion by choosing $T = 1$, which
corresponds to the 68\% C.L., i.e.\ the $1\sigma$ error band.
In Sec.~\ref{sec:results}, the uncertainty bands thus determined are
presented both for the $B$-hadron FFs and for the physical observables
evaluated with them.

%%%%%%%%%%%%%%%%%%%%%%%%%%%%%%%%%%%%%%%%%%%%%%%%%%%%%%%%%%%%%%%%%%%%%%%
\section{Results and discussion} \label{sec:results}
%%%%%%%%%%%%%%%%%%%%%%%%%%%%%%%%%%%%%%%%%%%%%%%%%%%%%%%%%%%%%%%%%%%%%%%

We are now in a position to present our results for the $B$-hadron FFs both at
NLO and NNLO and to compare the resulting theoretical predictions with the
experimental data fitted to, so as to check directly the consistency and
goodness of our fits.
We also compare our $B$-hadron FFs with the NLO ones presented by
Kniehl, Kramer, Schienbein and Spiesberger (KKSS) \cite{Kniehl:2008zza}.

In Table~\ref{Data-Set}, for each of the four experimental data sets $n$,
from ALEPH \cite{Heister:2001jg}, DELPHI \cite{DELPHI:2011aa}, OPAL
\cite{Abbiendi:2002vt}, and SLD \cite{Abe:2002iq}, the number
$N_n^{\mathrm{data}}$ of data points included in the NLO and NNLO fits and
the normalization factors ${\cal N}_n$ and the $\chi_n^2(p)$ values thus
obtained are specified together with the total number of data points and the
values of $\chi_{\mathrm{global}}^2(p)$ and
$\chi_{\mathrm{global}}^2(p)/\mathrm{d.o.f.}$
There are $59-3=56$ degrees of freedom.
The NLO and NNLO fits are both excellent, with
$\chi_{\mathrm{global}}^2(p)/\mathrm{d.o.f.}$ values of order unity.
As expected on general grounds, $\chi_{\mathrm{global}}^2(p)/\mathrm{d.o.f.}$ is
reduced as one passes from NLO to NNLO.
This is also true for the individual data sets, except for the most recent one,
from DELPHI.
The NLO and NLO fit results for $p$ are summarized in
Table.~\ref{tab:parameters}.

%%%%%%%%%%%%%%
% TABLE 1
%%%%%%%%%%%%%%
%--------------------------------
\begin{table}[b]
  \caption{\label{Data-Set}%
    Numbers $N_n^{\mathrm{data}}$ of data points from data set $n$ included
    in the NLO and NNLO fits and normalization factors ${\cal N}_n$ and
    $\chi_n^2$ values thus obtained; total number of data points;
    $\chi_{\mathrm{global}}^2(p)$ values; and
    $\chi_{\mathrm{global}}^2(p)/\mathrm{d.o.f.}$ values.} 
\begin{ruledtabular}
\tabcolsep=0.06cm \footnotesize
\begin{tabular}{lccccc}
  Collaboration & $N_n^{\mathrm{data}}$ & ${\cal N}_n^{\mathrm{NLO}}$ &
  ${\cal N}_n^{\mathrm{NNLO}}$ & $\chi_n^{2,\mathrm{NLO}}$ &
  $\chi_n^{2,\mathrm{NNLO}}$ \\  \hline
ALEPH~\cite{Heister:2001jg} & 18 & 1.0008 & 1.0011 & 14.376 & 12.269 \\
DELPHI~\cite{DELPHI:2011aa} &  8 & 0.9993 & 1.0058 & 7.535 & 15.377 \\
OPAL~\cite{Abbiendi:2002vt} & 15 & 0.9951 & 0.9958 & 35.594 & 20.002 \\
SLD~\cite{Abe:2002iq}       & 18 & 1.0030 & 0.9996 & 25.675 & 14.195 \\ 
\hline
Total                              & 59 & & & 83.180 & 61.844 \\ 
$\chi_{\mathrm{global}}^2(p)/\mathrm{d.o.f.}$ &    & & &  1.485 &  1.104 \\
\end{tabular}
\end{ruledtabular}
\end{table}
%--------------------------------

%%%%%%%%%%%%%%
% TABLE 2
%%%%%%%%%%%%%%
%--------------------------------
\begin{table*}[t]
  \caption{\label{tab:parameters}%
    Values of the fit parameters in Eq.~(\ref{input}) obtained at NLO and
    NNLO.}
\begin{ruledtabular}
\begin{tabular}{lccc}
Order & $N_b$ & $\alpha_b$ & $\beta_b$ \\ \hline
NLO  &$2575.014$  & $15.424$ & $2.394$  \\
NNLO & $1805.896$ & $ 14.168$ & $2.341$ \\
\end{tabular}
\end{ruledtabular}
\end{table*}
%--------------------------------

\begin{figure*}[htb]
  \includegraphics[width=\textwidth]{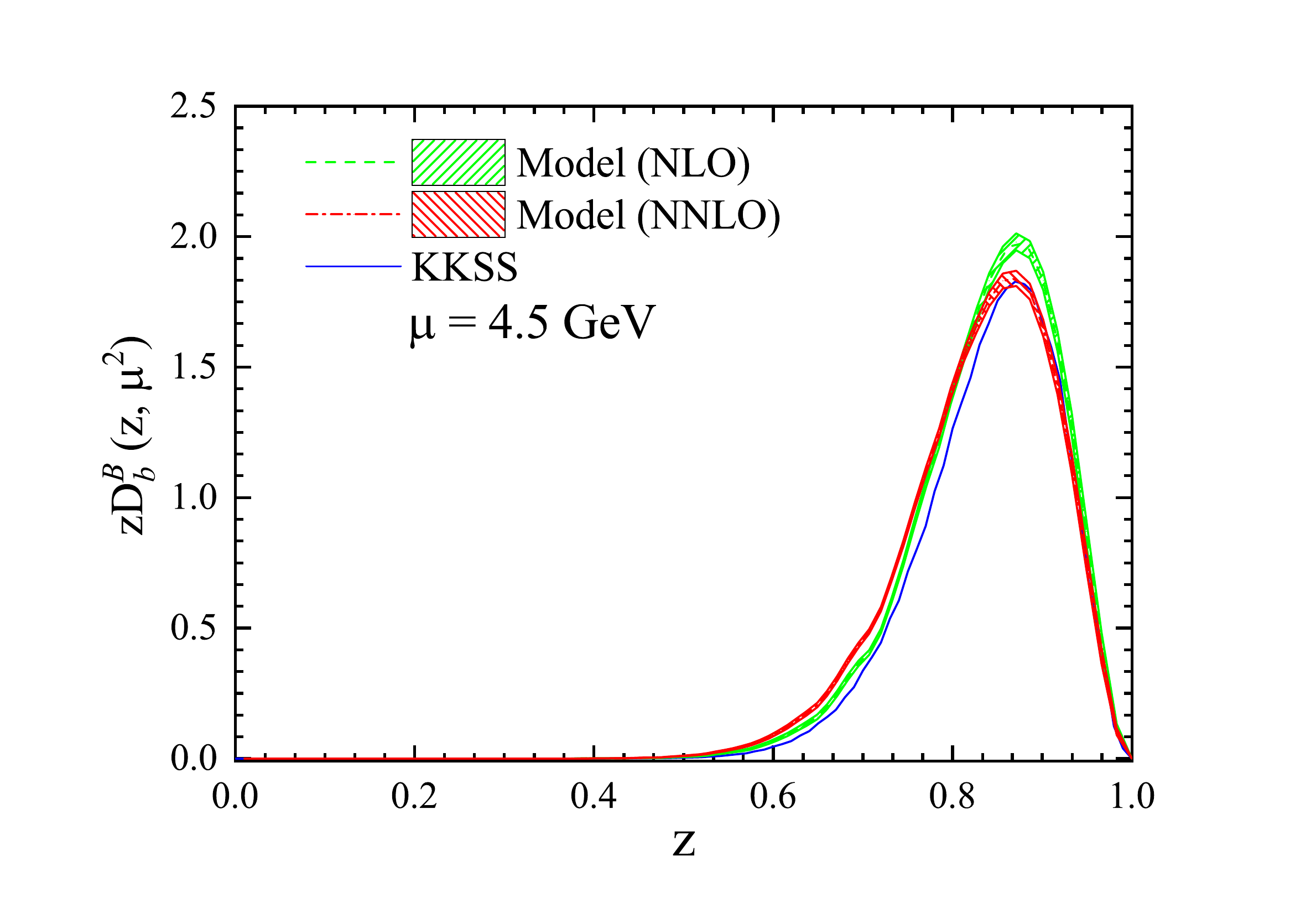}
 	\begin{center}
          \caption{\label{fig:bbar-NLO-NNLO-Q4.5}%
            Line shapes of $zD_b^B(z, \mu_{0})$ with $\mu_0=4.5$~GeV at NLO
            (green dashed line) and NNLO (red dot-dashed line) and their
            experimental uncertainty bands (green and red hatched areas).
            The KKSS result \cite{Kniehl:2008zza} (blue solid line) is shown
            for comparison.}
	\end{center}
\end{figure*}

In Fig.~\ref{fig:bbar-NLO-NNLO-Q4.5}, the $z$ distributions of the NLO and
NNLO $b\to B$ FFs at the initial scale $\mu_F=\mu_0$ are compared with each other.
The NLO and NNLO results agree in shape and position of maximum, but differ in
normalization.
This difference is induced by the $\mathcal{O}(\alpha_s^2)$ correction terms
in the hard-scattering cross sections and in the timelike splitting functions,
and it is compensated in the physical cross sections to be compared with the
experimental data up to terms beyond $\mathcal{O}(\alpha_s^2)$.
The error bands determined as described in Sec.~\ref{sec:errorcalculation} are
also shown in Fig.~\ref{fig:bbar-NLO-NNLO-Q4.5}.
They are dominated by the experimental errors, which explains why they are
not reduced by passing from NLO to NNLO.
In Fig.~\ref{fig:bbar-NLO-NNLO-Q4.5}, the KKSS $b\to B$ FF
\cite{Kniehl:2008zza} is included for comparison.
It somewhat undershoots our NLO $b\to B$ FF, which we attribute to the impact
of the DELPHI data \cite{DELPHI:2011aa}, which were not available at the time
of the analysis in Ref.~\cite{Kniehl:2008zza}.

%------------------------------------------------
%------------------------------------------------

\begin{figure*}[htb]
\includegraphics[width=0.480\textwidth]{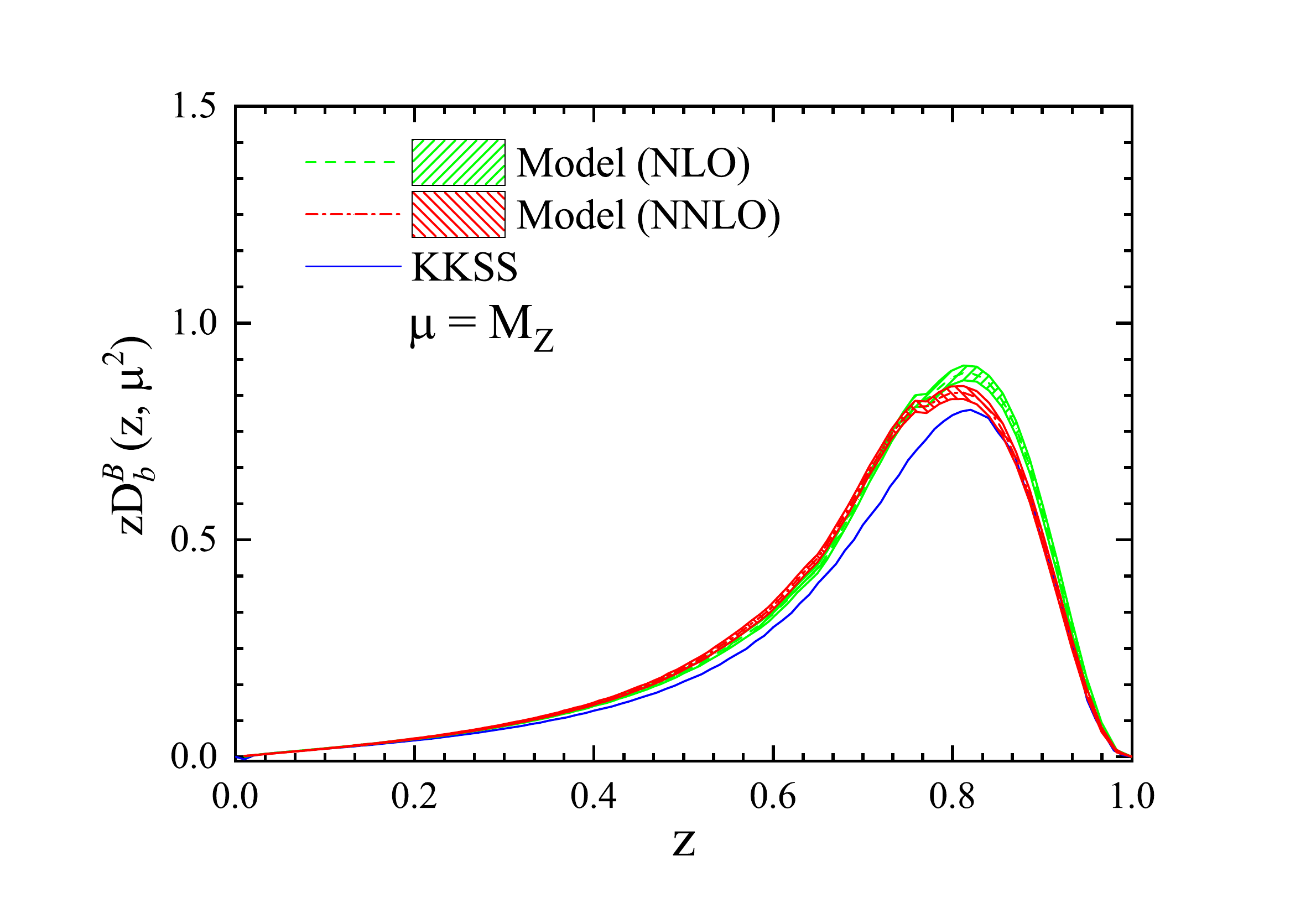}
\includegraphics[width=0.480\textwidth]{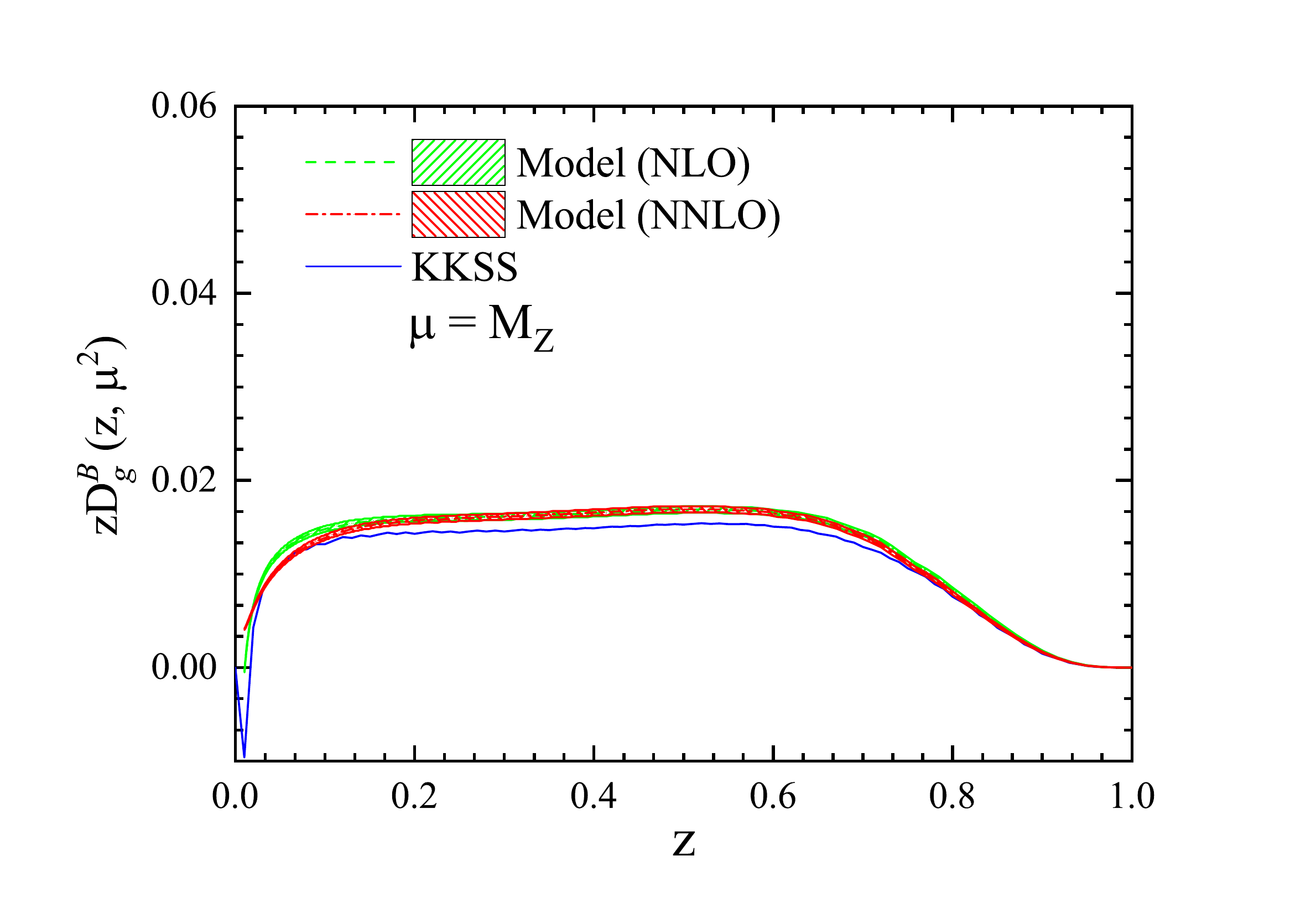}
\includegraphics[width=0.480\textwidth]{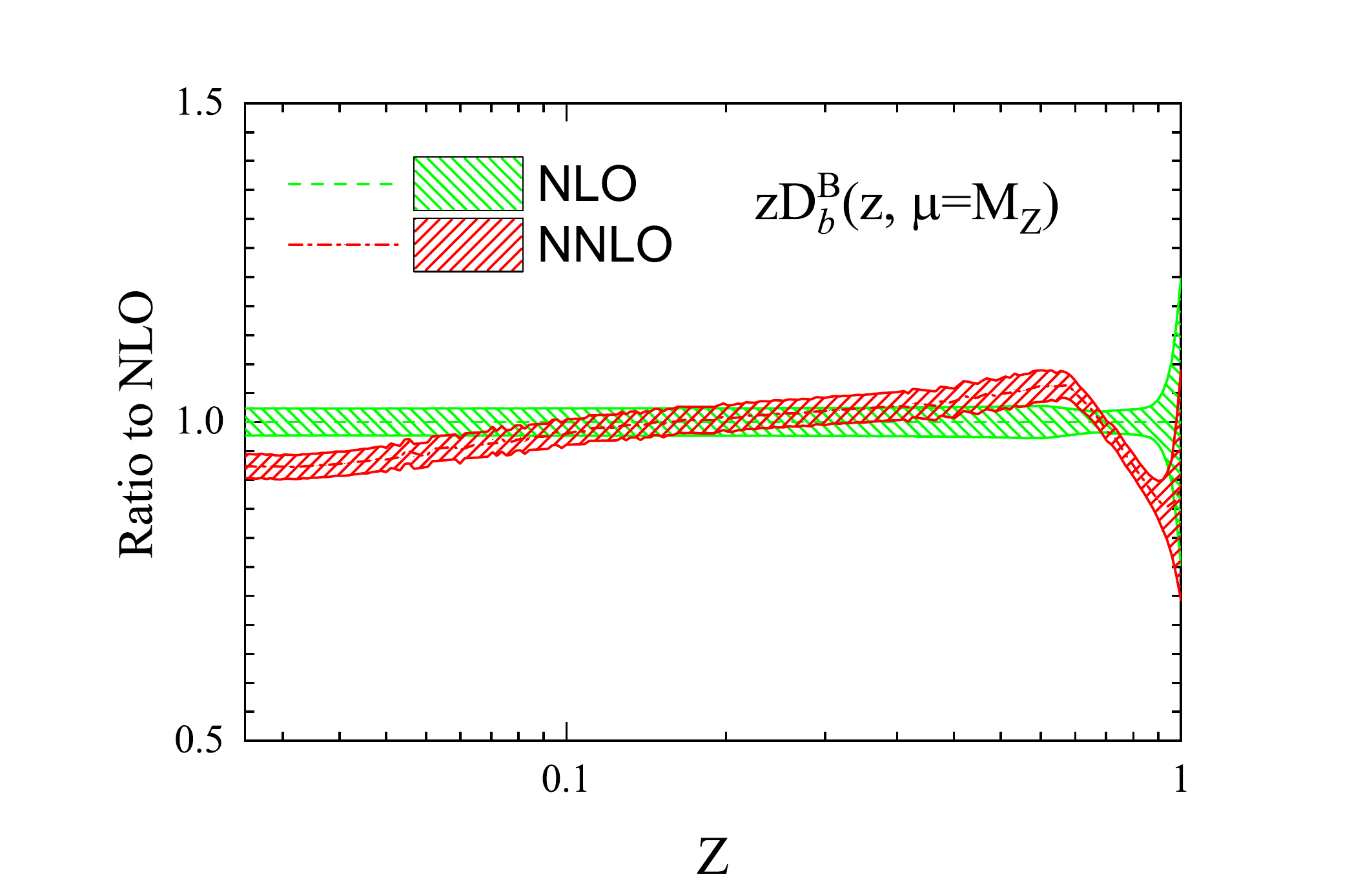}
\includegraphics[width=0.480\textwidth]{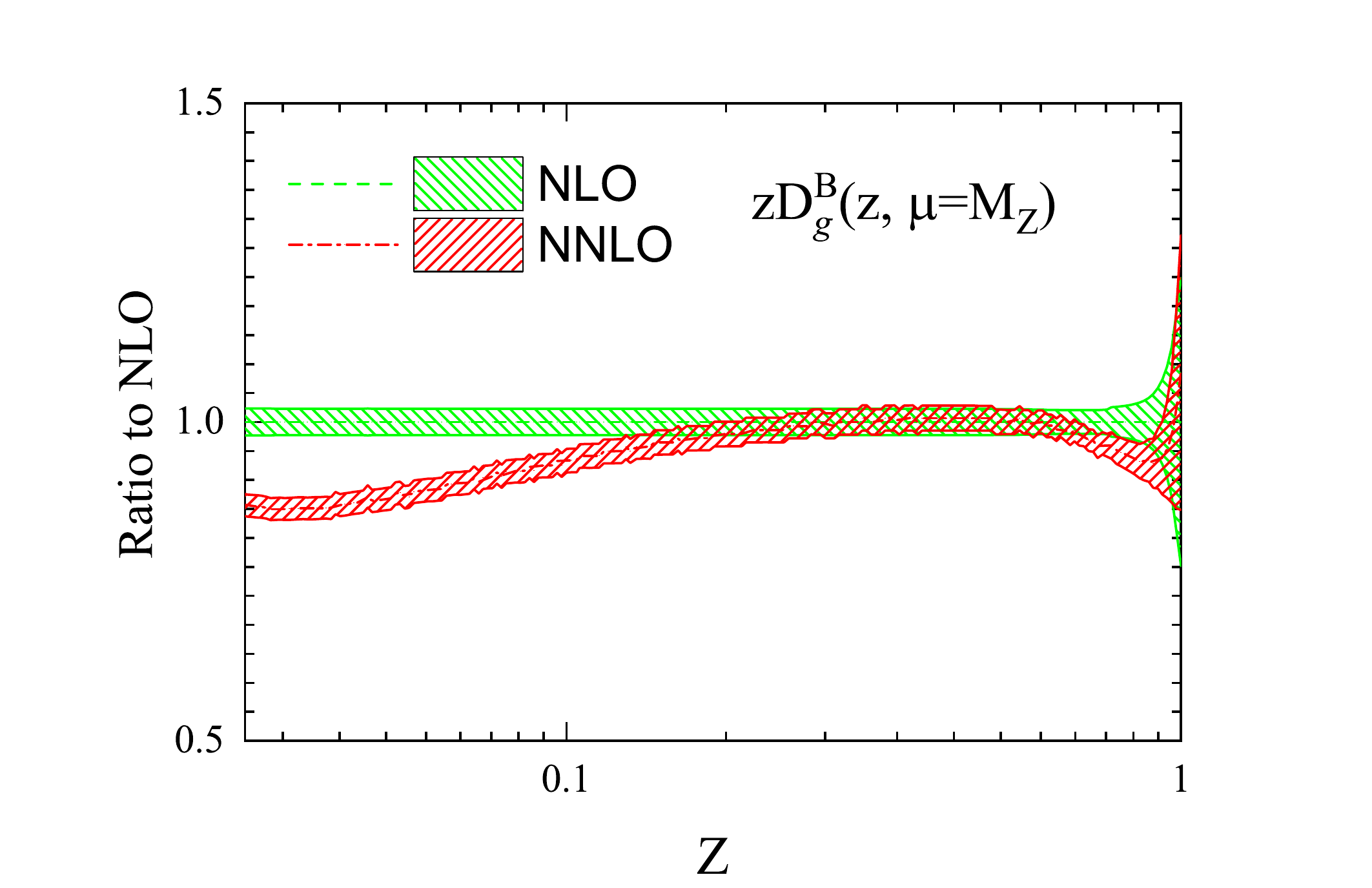}  
	\begin{center}
          \caption{\label{fig:bbar-NLO-NNLO-MZ}%
           Line shapes of (a) $zD_b^B(z,M_Z)$ and (b) $zD_g^B(z,M_Z)$ at NLO
            (green dashed lines) and NNLO (red dot-dashed lines) and their
            experimental uncertainty bands (green and red hatched areas).
            The KKSS results \cite{Kniehl:2008zza} (blue solid lines) are shown
            for comparison. NLO and NNLO results for (c) $zD_b^B(z,M_Z)$ and
            (d) $zD_g^B(z,M_Z)$ normalized with respect to the NLO central
            values.}
	\end{center}
\end{figure*}

In Fig.~\ref{fig:bbar-NLO-NNLO-MZ}(a), the analysis of
Fig.~\ref{fig:bbar-NLO-NNLO-Q4.5} is repeated for $\mu_F=M_Z$, the CM energy of
the experimental data fitted to.
Our NLO and NNLO $b\to B$ FFs are now closer together, the remaining
difference being entirely due to the $\mathcal{O}(\alpha_s^2)$ corrections to
the hard-scattering cross sections.
On the other hand, the difference between the NLO $b\to B$ FF and the KKSS one
is hardly affected by the DGLAP evolution from $\mu_0$ to $M_Z$, as it is due
to a difference in the collection of experimental data fitted to.
Figure~\ref{fig:bbar-NLO-NNLO-MZ}(b) is the counterpart of
Fig.~\ref{fig:bbar-NLO-NNLO-MZ}(a) for the $g\to B$ FF, which is generated
by DGLAP evolution from the initial condition $D_g^B(z,\mu_0)=0$, as explained
in Sec.~\ref{sec:framework}.
Our NLO and NNLO results are now very similar; the KKSS result again falls
below our NLO result.
The comparisons between our NLO and NNLO results for the $b\to B$ and $g\to B$
FFs are refined in Figs.~\ref{fig:bbar-NLO-NNLO-MZ}(c) and (d), respectively,
where these FFs and their error bands are normalized with respect to the
central values at NLO.
Deviations occur at small and large values of $z$, which are outside the focus
of our present study.
They are due to large soft-gluon and threshold logarithms, respectively, which
are included through $\mathcal{O}(\alpha_s^2)$ at NNLO, but only to
$\mathcal{O}(\alpha_s)$ at NLO.
These logarithms invalidate the fixed-order treatment at small and large values
of $z$ and should be resummed.
This is, however, beyond the scope of our present analysis and left for future
work.

%------------------------------------------------
%------------------------------------------------

\begin{figure*}[htb]
\includegraphics[width=\textwidth]{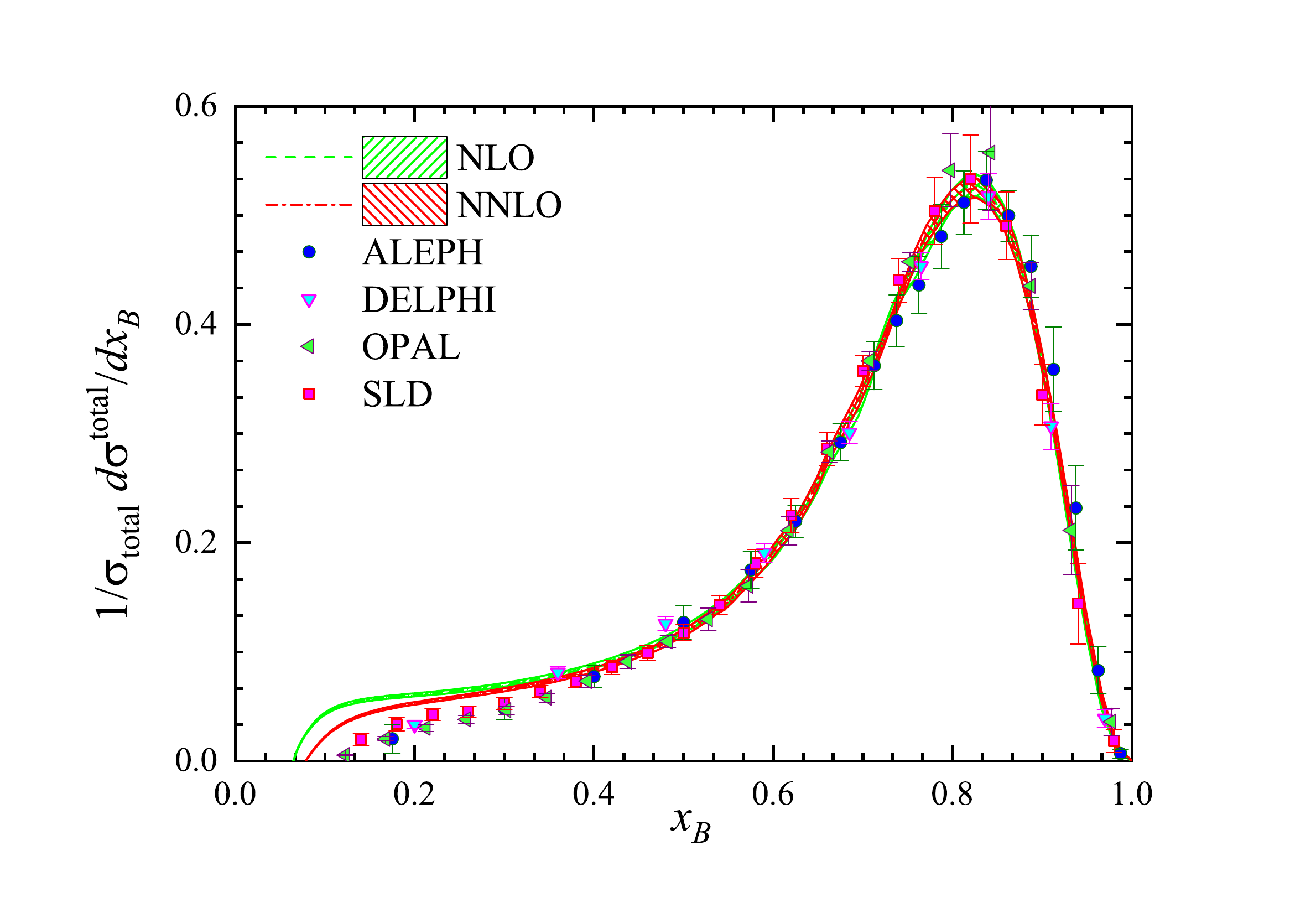}
\begin{center}
  \caption{\label{fig:ALEPH}%
    The NLO (green dashed line) and NNLO (red dot-dashed line) results for
    $(1/\sigma_{\mathrm{tot}})d\sigma(e^+e^-\to B+X)/dx_B$ evaluated with our
    respective $B$-hadron FF sets are compared with the experimental data
    fitted to, from
    ALEPH \cite{Heister:2001jg}, DELPHI \cite{DELPHI:2011aa},
    OPAL \cite{Abbiendi:2002vt}, and SLD \cite{Abe:2002iq}.
    The uncertainty bands (green and red hatched areas) stem from those of
    the $B$-hadron FFs.}
\end{center}
\end{figure*}

In Fig.~\ref{fig:ALEPH}, the NLO and NNLO results for
$(1/\sigma_{\mathrm{tot}})d\sigma(e^+e^-\to B+X)/dx_B$ evaluated with our
respective $B$-hadron FF sets are compared with the experimental data fitted to.
The uncertainty bands stem from those of the $B$-hadron FFs and are of
experimental origin.
We observe that the experimental data are in good mutual agreement and are
well described both by the NLO and NNLO results down to $x_B$ values of 0.4,
say, as for both line shape and normalization.
The NNLO description does somewhat better at lower values of $x_B$, which
explains the lower value of $\chi_{\mathrm{global}}^2(p)/\mathrm{d.o.f.}$ in
Table~\ref{Data-Set}.
The failure of the theoretical descriptions in the small-$x_B$ regime is, of
course, a direct consequence of the small-$x_B$ cuts applied.

%------------------------------------------------
%------------------------------------------------

\begin{figure*}[htb]
\includegraphics[width=0.480\textwidth]{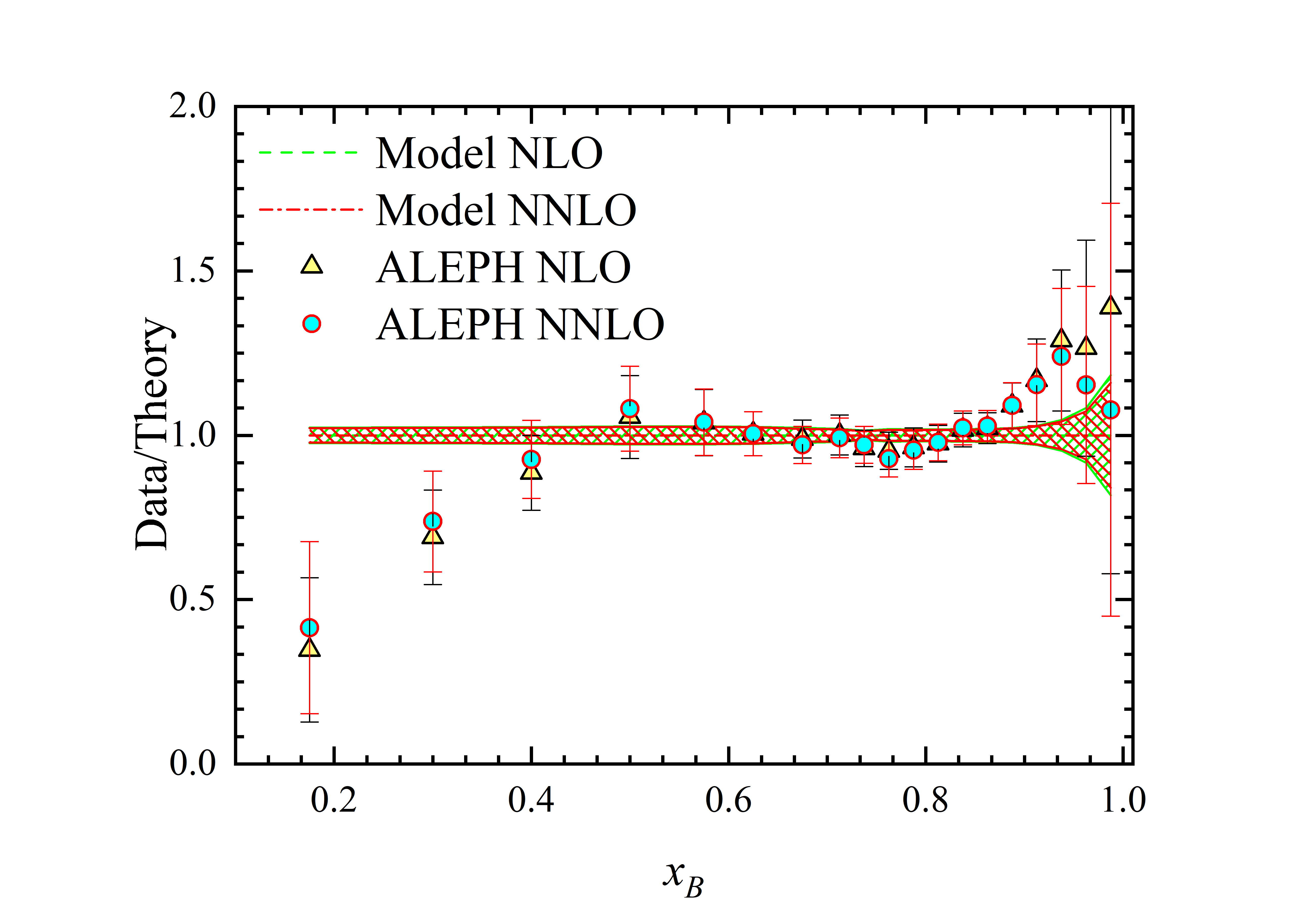}  	
\includegraphics[width=0.480\textwidth]{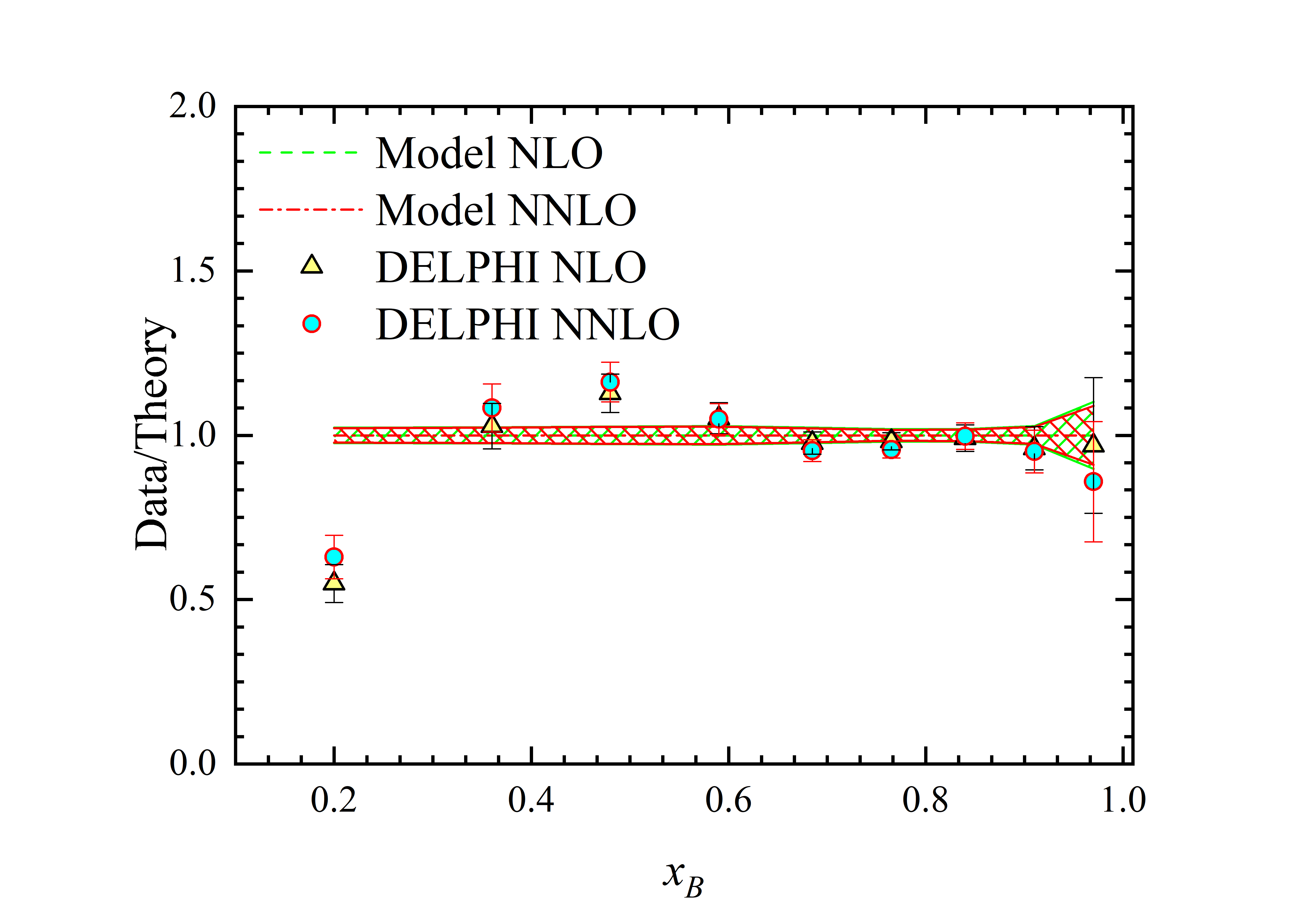}
\includegraphics[width=0.480\textwidth]{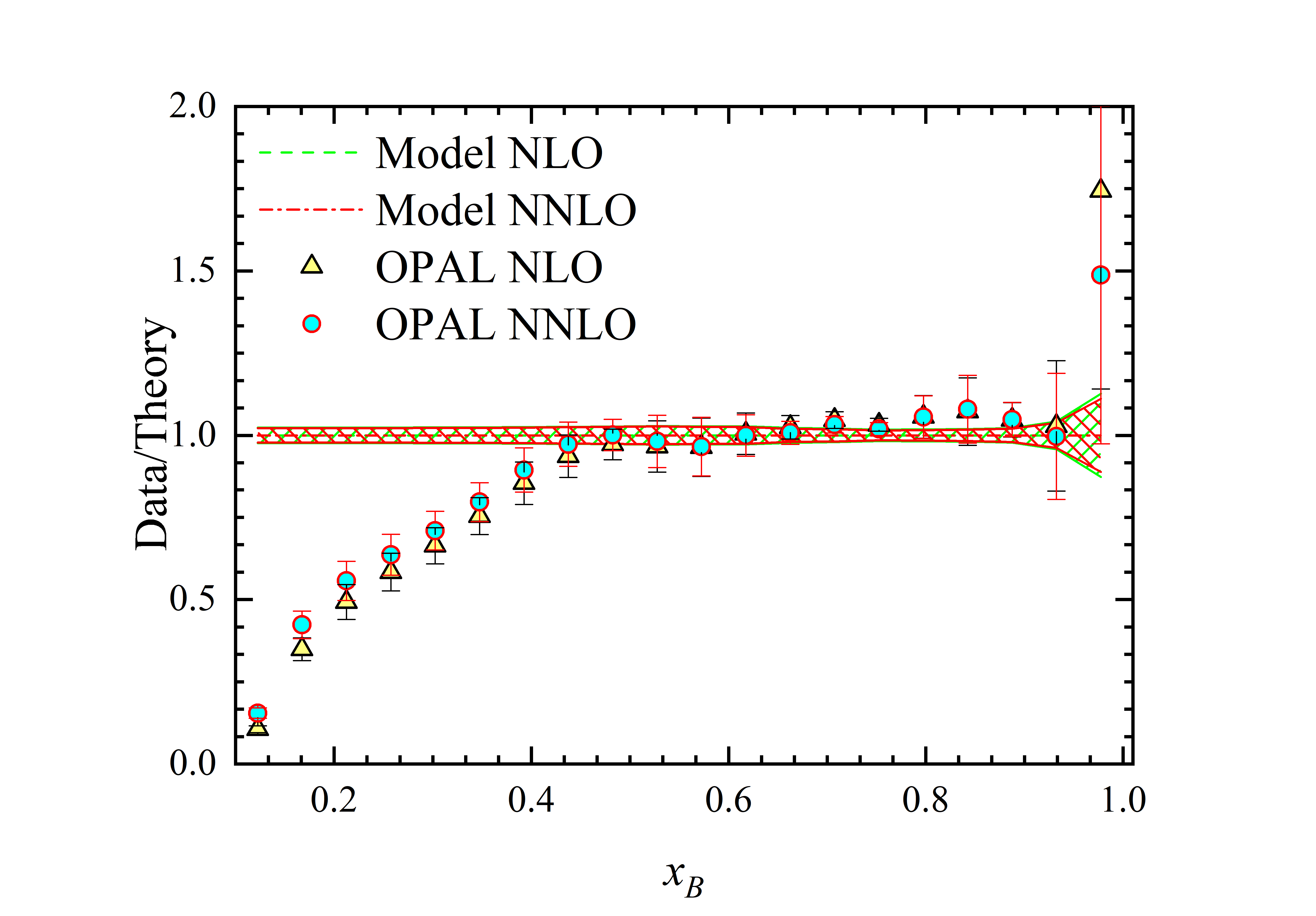}
\includegraphics[width=0.480\textwidth]{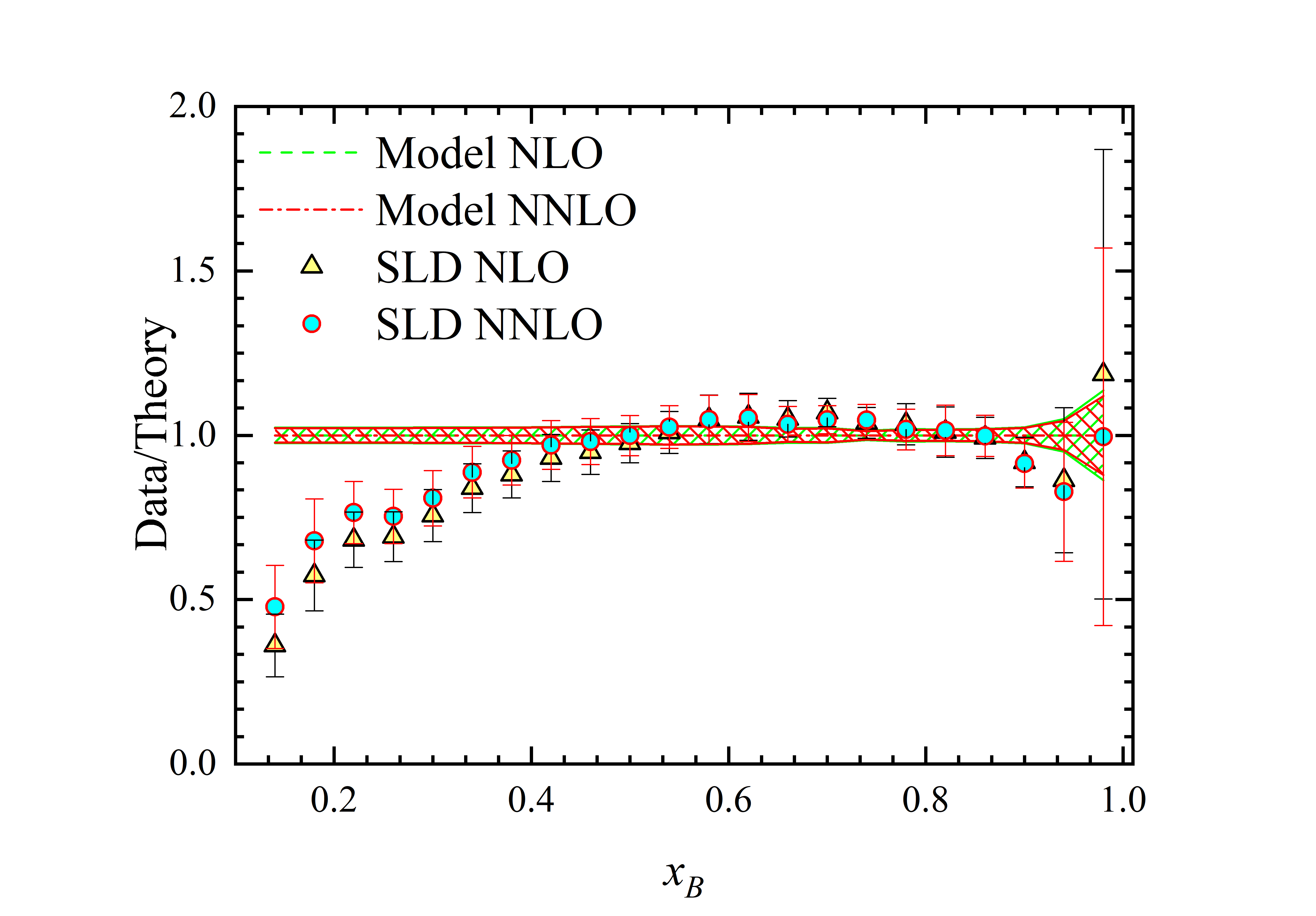}
\begin{center}
  \caption{\label{fig:ALEPH-Ratio}%
    (a) ALEPH \cite{Heister:2001jg}, (b) DELPHI \cite{DELPHI:2011aa},
    (c) OPAL \cite{Abbiendi:2002vt}, and (d) SLD \cite{Abe:2002iq} data of
    $(1/\sigma_{\mathrm{tot}})d\sigma(e^+e^-\to B+X)/dx_B$ normalized with
    respect to our NLO (green hatched bands) and NNLO (red hatched bands)
    results.}
\end{center}
\end{figure*}

For better visibility, we present the information contained in
Fig.~\ref{fig:ALEPH} as data over theory plots in
Fig.~\ref{fig:ALEPH-Ratio}, one for each experiment.
Specifically, the experimental data are in turn normalized to the NLO and
NNLO central values.
As already explained above, the NLO and NNLO uncertainty bands are very
similar.
As already visible in Fig.~\ref{fig:ALEPH}, the experimental data consistently
undershoot the NLO and NNLO results in the small-$x_B$ regime.
On the other hand, their large-$x_B$ behavior is nonuniform.
While the ALEPH and OPAL data overshoot the NLO and NNLO results in the
upper $x_B$ range, there is nice agreement for the DELPHI and SLD data.

%------------------------------------------------
%------------------------------------------------

%
%%%%%%%%%%%%%%%%%%%%%%%%%%%%%%%%%%%%%%%%%%%%%%%%%%%%%%%%%%%%%%%%%%%%%%%
\section{$B$-hadron production by top-quark decay} \label{sec:B-meson-LHC}
%%%%%%%%%%%%%%%%%%%%%%%%%%%%%%%%%%%%%%%%%%%%%%%%%%%%%%%%%%%%%%%%%%%%%%%
%

As a topical application of our $B$-hadron FFs, we study inclusive single
$B$-hadron production at the LHC.
$B$ hadrons may be produced directly or through the decay of heavier particles,
including the $Z$ boson, the Higgs boson, and the top quark.
For definiteness, we concentrate here on the latter process, $t\to BW^++X$,
where $X$ collectively denotes any other final-state particles.
This allows one to study properties of the top quark, such as its degree of
polarization in a given production mode, which includes single and pair
production.
We thus consider both unpolarized and polarized top quarks.

We work in the rest frame of the top quark.
The partial width of the decay $t\to BW^++X$, differential in the scaled
$B$-hadron energy $x_B$ and the angle $\theta_P$ enclosed between the top-quark
polarization three-vector $\vec{P}$ and the $B$-hadron three-momentum
$\vec{p}_B$ is given by 
\begin{equation}\label{eq:pol}
  \frac{d^2\Gamma}{dx_B\,d\cos\theta_P}(t\to BW^++X)
  =\frac{1}{2}\left(\frac{d\Gamma^{\mathrm{unpol}}}{dx_B}
  +P\frac{d\Gamma^{\mathrm{pol}}}{dx_B}\cos\theta_P\right),
\end{equation}
where $P=|\vec{P}\,|$ is the degree of polarization.
In the ZM-VFNS, we have
%--------------------------------
\begin{equation}\label{eq:master}
  \frac{d\Gamma^{\mathrm{unpol/pol}}}{dx_B}
  =\sum_{i=b,g}\int_{x_i^{min}}^{x_i^{\rm max}}\frac{dx_i}{x_i}\,
  \frac{d\Gamma_i^{\mathrm{unpol/pol}}}{dx_i}(x_i,\mu_R,\mu_F)
  D_i^{B}\left(\frac{x_B}{x_i},\mu_F\right), 
\end{equation}
%--------------------------------
where $d\Gamma_i^{\mathrm{unpol}}/dx_i$ and $d\Gamma_i^{\mathrm{pol}}/dx_i$ refer to
the parton-level decay $t \to iW^++X$, differential in the scaled energy $x_i$
of parton $i=b,g$.
In the top-quark rest frame, we have $x_B = E_B/E_b^{\mathrm{max}}$ and
$x_i = E_i/E_b^{\mathrm{max}}$, where $E_B$ and $E_i$ are the energies of the $B$
hadron and parton $i$, and $E_b^{\mathrm{max}}$ is the maximum energy of the
bottom quark.
In our application of the ZM-VFNS, where $m_b\ll\mu_F=\mathcal{O}(m_t)$, the
bottom quark is taken to be massless.
By the same token, we also neglect the $B$-hadron mass $m_B$. 
So far, $d\Gamma_i^{\mathrm{unpol}}/dx_i$ and $d\Gamma_i^{\mathrm{pol}}/dx_i$ are
only available through NLO;
analytic expressions may be found in
Refs.~\cite{Corcella:2001hz,Cacciari:2002re,Kniehl:2012mn} and
Refs.~\cite{Fischer:1998gsa,Fischer:2001gp,Nejad:2013fba,Nejad:2014sla},
respectively.
In Ref.~\cite{Nejad:2014sla}, $\theta_P$ is taken to be enclosed between
$\vec{P}$ and the $W$-boson three-momentum $\vec{p}_W$.
Although a consistent analysis is presently limited to NLO, we also employ our
NNLO $B$-hadron FF set to explore the possible size of the NNLO corrections.

\begin{figure*}[htb]
\includegraphics[width=0.480\textwidth]{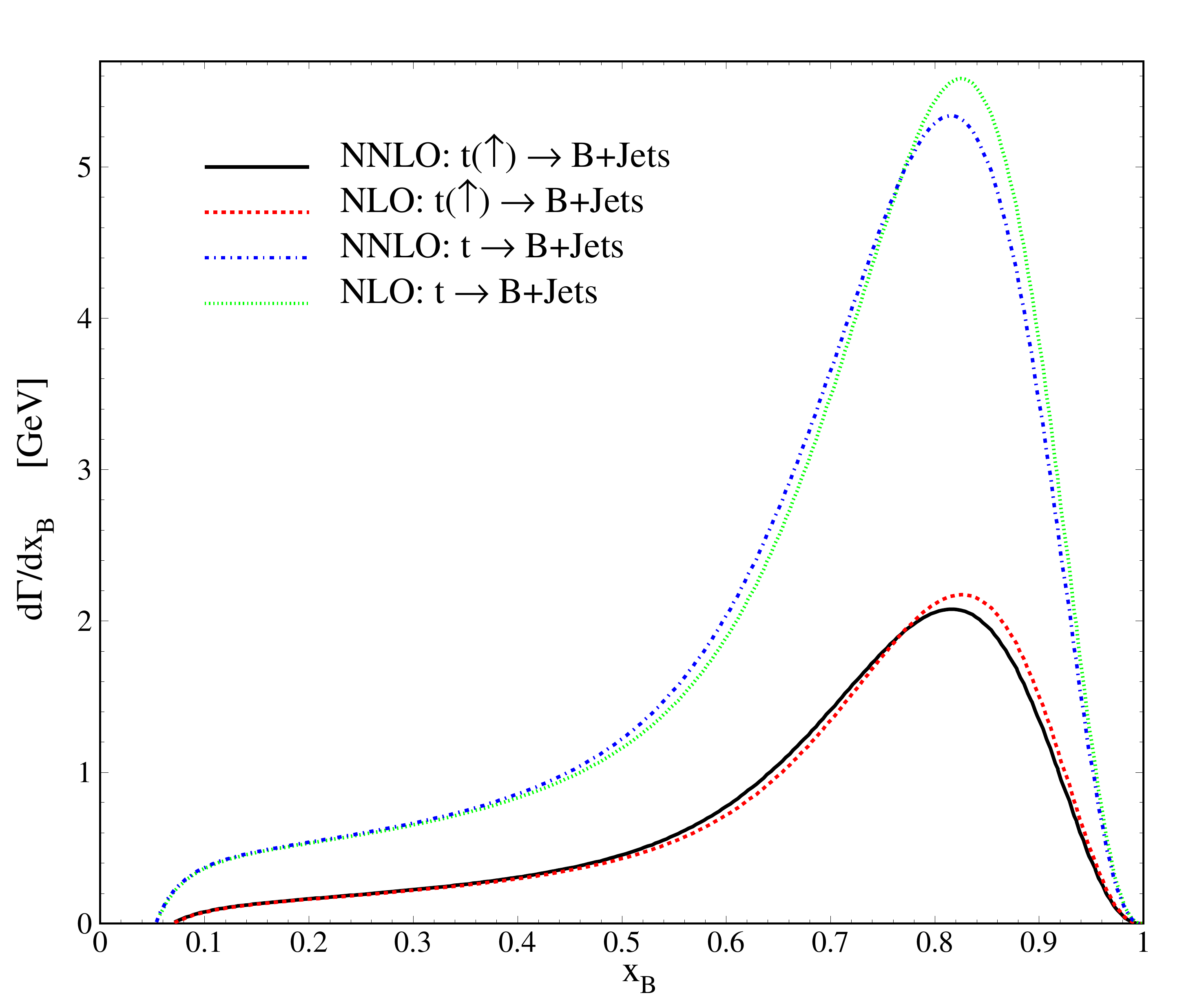}
\includegraphics[width=0.480\textwidth]{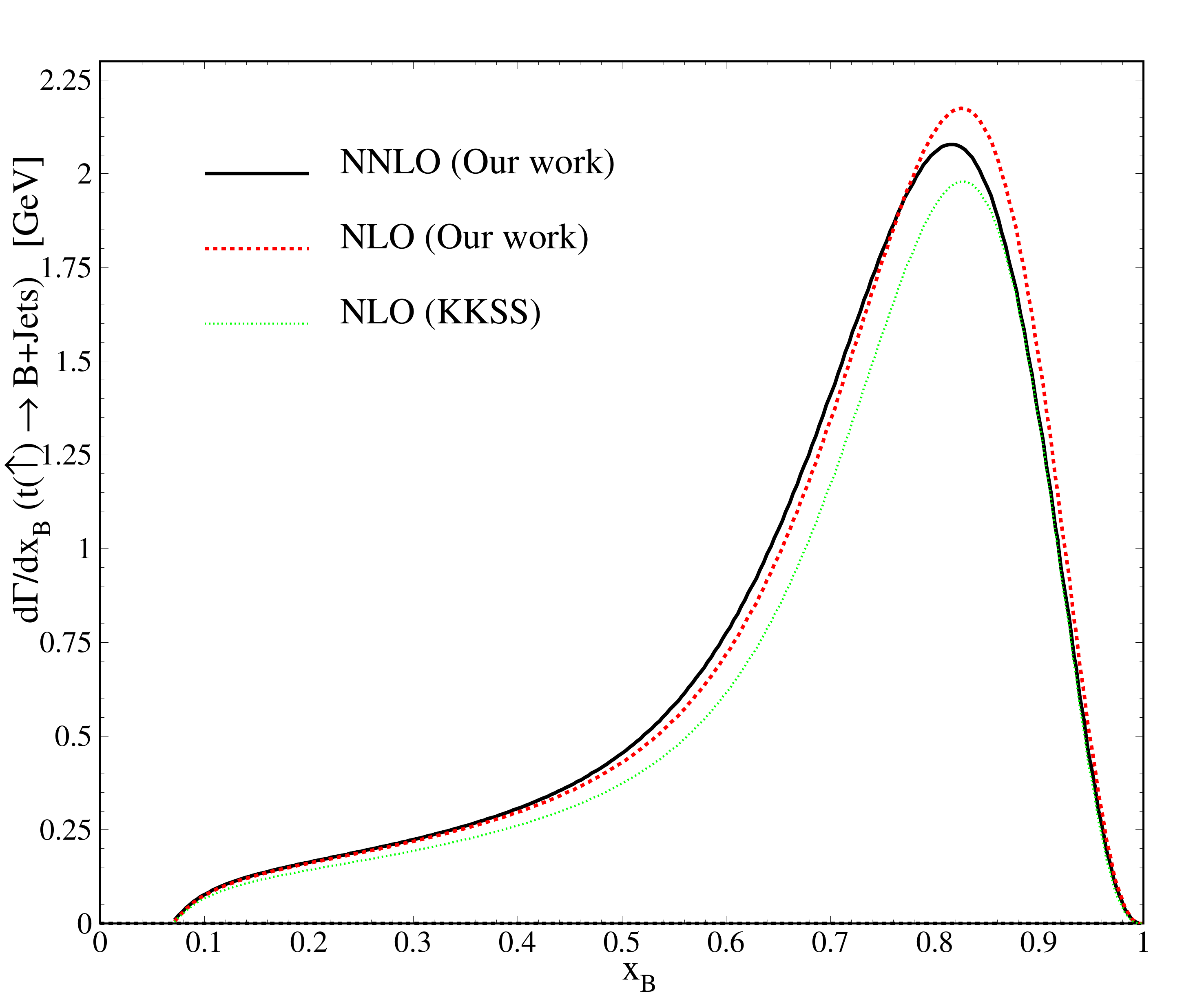}
\begin{center}
  \caption{\label{top}%
    (a) NLO predictions of $d\Gamma^{\mathrm{unpol}}/dx_B$ (green dotted line) and
    $d\Gamma^{\mathrm{pol}}/dx_B$ (red dashed line), evaluated with our NLO
    $B$-hadron FF set.
    For comparison, the evaluations with our NNLO $B$-hadron FF set are also
    included (blue dot-dashed and black solid lines).
    (b) The results for $d\Gamma^{\mathrm{pol}}/dx_B$ in Fig.~\ref{top}(a) are
    compared to the evaluation with the KKSS $B$-hadron FF set
    \cite{Kniehl:2008zza} (green dotted line).}
\end{center}
\end{figure*}

In our numerical analysis, we use $m_b=4.5$~GeV, $m_W=80.379$~GeV, and 
$m_t = 173.0$~GeV \cite{Tanabashi:2018oca}, and choose $\mu_R = \mu_F = m_t$.
In Fig.~\ref{top}(a), we present the NLO predictions of
$d\Gamma^{\mathrm{unpol}}/dx_B$ and $d\Gamma^{\mathrm{pol}}/dx_B$, evaluated with our
NLO $B$-hadron FF set.
For comparison, the evaluations with our NNLO $B$-hadron FF set are also
included.
We observe from Fig.~\ref{top}(a) that switching from the NLO $B$-hadron FF set
to the NNLO one slightly smoothens the theoretical prediction, decreasing it in
the peak region and increasing it in the tail region thereunder.
At the same time, the peak position is shifted towards smaller values of $x_B$.
The change in normalization is of order 5\% at most.
These effects should mark an upper limit of the total NNLO corrections because
the as-yet-unknown NNLO corrections to $d\Gamma_i^{\mathrm{unpol}}/dx_i$ and
$d\Gamma_i^{\mathrm{pol}}/dx_i$ are expected to give rise to some compensation if
FF universality is realized in nature.
In Fig.~\ref{top}(b), the results for $d\Gamma^{\mathrm{pol}}/dx_B$ in
Fig.~\ref{top}(a) are compared to the evaluation with the KKSS $B$-hadron FF
set \cite{Kniehl:2008zza}.
As in Figs.~\ref{fig:bbar-NLO-NNLO-Q4.5}, \ref{fig:bbar-NLO-NNLO-MZ}(a), and
  \ref{fig:bbar-NLO-NNLO-MZ}(b), the NLO result is somewhat reduced by
switching to the KKSS $B$-hadron FF set. 

%------------------------------------------------
%------------------------------------------------

%
%%%%%%%%%%%%%%%%%%%%%%%%%%%%%%%%%%%%%%%%%%%%%%%%%%%%%%%%%%%%%%%%%%%%%%%
\section{Summary and Conclusions} \label{sec:conclusion}
%%%%%%%%%%%%%%%%%%%%%%%%%%%%%%%%%%%%%%%%%%%%%%%%%%%%%%%%%%%%%%%%%%%%%%%
%
In this paper, we determined nonperturbative FFs for $B$ hadrons, both at NLO
and NNLO in the ZM-FVNS, by fitting to all available experimental data of
inclusive single $B$-hadron production in $e^+e^-$ annihilation,
$e^+e^-\to B+X$, from ALEPH \cite{Heister:2001jg}, DELPHI \cite{DELPHI:2011aa},
OPAL \cite{Abbiendi:2002vt}, and SLD \cite{Abe:2002iq}.
We then applied these $B$-hadron FFs to provide NLO predictions for inclusive
$B$-hadron production by top-quark decay, $t\to BW^++X$, both for unpolarized
and polarized top quarks.

Our analysis updates and improves similar ones in the literature 
\cite{Binnewies:1998vm,Kniehl:2008zza} in the following respects.
We included the DELPHI data \cite{DELPHI:2011aa}, which had not been available
then.
For the first time, we advanced to NNLO in a fit of $B$-hadron FFs.
We performed a careful estimation of the experimental uncertainties in our
$B$-hadron FFs using the Hessian approach.

We adopted the simple power ansatz of Eq.~(\ref{input}) and obtained for the
three fit parameters appearing therein the values listed in
Table~\ref{tab:parameters}.
The goodness of the NLO and NNLO fits turned out to be excellent, with
$\chi^2/\mathrm{d.o.f.}$ values of 1.485 and 1.104, respectively
(see Table~\ref{Data-Set}).
As expected on general grounds, the fit quality is improved by ascending to
higher orders of perturbation theory.

We encourage the LHC Collaborations to measure the $x_B$ distribution of the
partial width of the decay $t\to BW^++X$, for two reasons.
On the one hand, this will allow for an independent determination of the
$B$-hadron FFs and thus provide a unique chance to test their universality and
DGLAP scaling violations, two important pillars of the QCD-improved parton
model of QCD.
On the other hand, this will allow for a determination of the top-quark
polarization, which should depend on the production mode.

The theoretical framework provided by the ZM-FVNS was quite appropriate for
the present analysis, since the characteristic energy scales of the considered
processes, $M_Z$ and $m_t$, greatly exceeded the bottom-quark mass $m_b$,
which could thus be neglected.
Possible theoretical improvements include the inclusion of finite-$m_b$ and
finite-$m_B$ effects, and the resummation of soft-gluon logarithms, which
extend the validity towards small values of $x_B$, and the resummation of
threshold logarithms, which extends the validity towards large values of $x_B$.
The general-mass variable-flavor-number scheme (GM-VFNS)
\cite{Kramer:2001gd,Kniehl:2004fy,Kniehl:2005mk,Kniehl:2005ej} provides a
consistent and natural finite-$m_b$ generalization of the ZM-VFNS on the basis
of the $\overline{\mathrm{MS}}$ factorization scheme \cite{Collins:1998rz}.
The processes considered here, $e^+e^-\to B+X$ \cite{Kneesch:2007ey},
$t\to BW^++X$ \cite{Kniehl:2012mn}, and $t(\uparrow)\to BW^++X$
\cite{Nejad:2016epx}, have all been worked out in the GM-VFNS at NLO, but not
yet at NNLO.
Finite-$m_B$ effects may be conveniently incorporated using the approach of
Refs.~\cite{Albino:2005gd,Albino:2006wz,Albino:2008fy}.
The implementation of such theoretical improvements reaches beyond the scope of
the present analysis and is left for future work.

%
%%%%%%%%%%%%%%%%%%%%%%%%%%%%%%%%%%%%%%%%%%%%%%%%%%%%%%%%%%%%%%%%%%%%%%%
\begin{acknowledgments}
%%%%%%%%%%%%%%%%%%%%%%%%%%%%%%%%%%%%%%%%%%%%%%%%%%%%%%%%%%%%%%%%%%%%%%%
%

Hamzeh Khanpour and Maryam Soleymaninia thank the School of Particles and 
Accelerators at the Institute for Research in Fundamental Sciences (IPM) for
financial support.
Hamzeh Khanpour is also grateful to the University of Science and Technology
of Mazandaran for financial support.
This work was supported in part by the German Federal Ministry for Education
and Research (BMBF) through Grant No.\ 05H18GUCC1.

\end{acknowledgments}
%

%\clearpage

%%%%%%%%%%%%%%%%%%%%%%%%%%%%%%%%%%%%%%%%%%%%%%%%%%%%%%%%%%%%%%%%%%%%%%%


\begin{thebibliography}{}

%\cite{Albajar:1988th}
\bibitem{Albajar:1988th} 
  C.~Albajar {\it et al.} [UA1 Collaboration],
  %``Measurement of the Bottom Quark Production Cross-Section in Proton - anti-Proton Collisions at s**(1/2) = 0.63-TeV,''
  Phys.\ Lett.\ B {\bf 213}, 405 (1988).
  doi:10.1016/0370-2693(88)91785-6
  %%CITATION = doi:10.1016/0370-2693(88)91785-6;%%
  %210 citations counted in INSPIRE as of 08 Apr 2019

%\cite{Gribov:1972ri}
\bibitem{Gribov:1972ri} 
  V.~N.~Gribov and L.~N.~Lipatov,
  %``Deep inelastic e p scattering in perturbation theory,''
  Sov.\ J.\ Nucl.\ Phys.\  {\bf 15}, 438 (1972)
  [Yad.\ Fiz.\  {\bf 15}, 781 (1972)].
  %%CITATION = SJNCA,15,438;%%
  %4003 citations counted in INSPIRE as of 08 Apr 2019

%\cite{Altarelli:1977zs}
\bibitem{Altarelli:1977zs} 
  G.~Altarelli and G.~Parisi,
  %``Asymptotic Freedom in Parton Language,''
  Nucl.\ Phys.\ B {\bf 126}, 298 (1977).
  doi:10.1016/0550-3213(77)90384-4
  %%CITATION = doi:10.1016/0550-3213(77)90384-4;%%
  %6586 citations counted in INSPIRE as of 08 Apr 2019

%\cite{Dokshitzer:1977sg}
\bibitem{Dokshitzer:1977sg} 
  Y.~L.~Dokshitzer,
  %``Calculation of the Structure Functions for Deep Inelastic Scattering and e+ e- Annihilation by Perturbation Theory in Quantum Chromodynamics.,''
  Sov.\ Phys.\ JETP {\bf 46}, 641 (1977)
  [Zh.\ Eksp.\ Teor.\ Fiz.\  {\bf 73}, 1216 (1977)].
  %%CITATION = SPHJA,46,641;%%
  %3609 citations counted in INSPIRE as of 08 Apr 2019

%\cite{Mitov:2006ic}
\bibitem{Mitov:2006ic} 
  A.~Mitov, S.~Moch and A.~Vogt,
  %``Next-to-Next-to-Leading Order Evolution of Non-Singlet Fragmentation Functions,''
  Phys.\ Lett.\ B {\bf 638}, 61 (2006)
  doi:10.1016/j.physletb.2006.05.005
  [hep-ph/0604053].
  %%CITATION = doi:10.1016/j.physletb.2006.05.005;%%
  %88 citations counted in INSPIRE as of 08 Apr 2019

%\cite{Moch:2007tx}
\bibitem{Moch:2007tx} 
  S.~Moch and A.~Vogt,
  %``On third-order timelike splitting functions and top-mediated Higgs decay into hadrons,''
  Phys.\ Lett.\ B {\bf 659}, 290 (2008)
  doi:10.1016/j.physletb.2007.10.069
  [arXiv:0709.3899 [hep-ph]].
  %%CITATION = doi:10.1016/j.physletb.2007.10.069;%%
  %60 citations counted in INSPIRE as of 08 Apr 2019

%\cite{Almasy:2011eq}
\bibitem{Almasy:2011eq} 
  A.~A.~Almasy, S.~Moch and A.~Vogt,
  %``On the Next-to-Next-to-Leading Order Evolution of Flavour-Singlet Fragmentation Functions,''
  Nucl.\ Phys.\ B {\bf 854}, 133 (2012)
  doi:10.1016/j.nuclphysb.2011.08.028
  [arXiv:1107.2263 [hep-ph]].
  %%CITATION = doi:10.1016/j.nuclphysb.2011.08.028;%%
  %49 citations counted in INSPIRE as of 08 Apr 2019

%\cite{Rijken:1996vr}
\bibitem{Rijken:1996vr} 
  P.~J.~Rijken and W.~L.~van Neerven,
  %``O (alpha-s**2) contributions to the longitudinal fragmentation function in e+ e- annihilation,''
  Phys.\ Lett.\ B {\bf 386}, 422 (1996)
  doi:10.1016/0370-2693(96)00898-2
  [hep-ph/9604436].
  %%CITATION = doi:10.1016/0370-2693(96)00898-2;%%
  %88 citations counted in INSPIRE as of 08 Apr 2019

%\cite{Rijken:1996npa}
\bibitem{Rijken:1996npa} 
  P.~J.~Rijken and W.~L.~van Neerven,
  %``O (alpha-s**2) contributions to the asymmetric fragmentation function in e+ e- annihilation,''
  Phys.\ Lett.\ B {\bf 392}, 207 (1997)
  doi:10.1016/S0370-2693(96)01529-8
  [hep-ph/9609379].
  %%CITATION = doi:10.1016/S0370-2693(96)01529-8;%%
  %53 citations counted in INSPIRE as of 08 Apr 2019

%\cite{Rijken:1996ns}
\bibitem{Rijken:1996ns} 
  P.~J.~Rijken and W.~L.~van Neerven,
  %``Higher order QCD corrections to the transverse and longitudinal fragmentation functions in electron - positron annihilation,''
  Nucl.\ Phys.\ B {\bf 487}, 233 (1997)
  doi:10.1016/S0550-3213(96)00669-4
  [hep-ph/9609377].
  %%CITATION = doi:10.1016/S0550-3213(96)00669-4;%%
  %105 citations counted in INSPIRE as of 08 Apr 2019

%\cite{Mitov:2006wy}
\bibitem{Mitov:2006wy} 
  A.~Mitov and S.~O.~Moch,
  %``QCD Corrections to Semi-Inclusive Hadron Production in Electron-Positron Annihilation at Two Loops,''
  Nucl.\ Phys.\ B {\bf 751}, 18 (2006)
  doi:10.1016/j.nuclphysb.2006.05.018
  [hep-ph/0604160].
  %%CITATION = doi:10.1016/j.nuclphysb.2006.05.018;%%
  %63 citations counted in INSPIRE as of 08 Apr 2019

%\cite{Anderle:2015lqa}
\bibitem{Anderle:2015lqa} 
  D.~P.~Anderle, F.~Ringer and M.~Stratmann,
  %``Fragmentation Functions at Next-to-Next-to-Leading Order Accuracy,''
  Phys.\ Rev.\ D {\bf 92}, no. 11, 114017 (2015)
  doi:10.1103/PhysRevD.92.114017
  [arXiv:1510.05845 [hep-ph]].
  %%CITATION = doi:10.1103/PhysRevD.92.114017;%%
  %38 citations counted in INSPIRE as of 08 Apr 2019

%\cite{Bertone:2017tyb}
\bibitem{Bertone:2017tyb} 
  V.~Bertone {\it et al.} [NNPDF Collaboration],
  %``A determination of the fragmentation functions of pions, kaons, and protons with faithful uncertainties,''
  Eur.\ Phys.\ J.\ C {\bf 77}, no. 8, 516 (2017)
  doi:10.1140/epjc/s10052-017-5088-y
  [arXiv:1706.07049 [hep-ph]].
  %%CITATION = doi:10.1140/epjc/s10052-017-5088-y;%%
  %34 citations counted in INSPIRE as of 08 Apr 2019

%\cite{Soleymaninia:2018uiv}
\bibitem{Soleymaninia:2018uiv} 
  M.~Soleymaninia, M.~Goharipour and H.~Khanpour,
  %``First QCD analysis of charged hadron fragmentation functions and their uncertainties at next-to-next-to-leading order,''
  Phys.\ Rev.\ D {\bf 98}, no. 7, 074002 (2018)
  doi:10.1103/PhysRevD.98.074002
  [arXiv:1805.04847 [hep-ph]].
  %%CITATION = doi:10.1103/PhysRevD.98.074002;%%
  %4 citations counted in INSPIRE as of 08 Apr 2019

%\cite{Soleymaninia:2019sjo}
\bibitem{Soleymaninia:2019sjo} 
  M.~Soleymaninia, M.~Goharipour and H.~Khanpour,
  %``Impact of unidentified light charged hadron data on the determination of pion fragmentation functions,''
  Phys.\ Rev.\ D {\bf 99}, no. 3, 034024 (2019)
  doi:10.1103/PhysRevD.99.034024
  [arXiv:1901.01120 [hep-ph]].
  %%CITATION = doi:10.1103/PhysRevD.99.034024;%%

%\cite{Soleymaninia:2017xhc}
\bibitem{Soleymaninia:2017xhc} 
  M.~Soleymaninia, H.~Khanpour and S.~M.~Moosavi Nejad,
  %``First determination of $D^{*+}$-meson fragmentation functions and their uncertainties at next-to-next-to-leading order,''
  Phys.\ Rev.\ D {\bf 97}, no. 7, 074014 (2018)
  doi:10.1103/PhysRevD.97.074014
  [arXiv:1711.11344 [hep-ph]].
  %%CITATION = doi:10.1103/PhysRevD.97.074014;%%
  %13 citations counted in INSPIRE as of 08 Apr 2019

%\cite{Binnewies:1998vm}
\bibitem{Binnewies:1998vm} 
  J.~Binnewies, B.~A.~Kniehl and G.~Kramer,
  %``Inclusive $B$ meson production in $e^{+} e^{-}$ and $p \bar{p}$ collisions,''
  Phys.\ Rev.\ D {\bf 58}, 034016 (1998)
  doi:10.1103/PhysRevD.58.034016
  [hep-ph/9802231].
  %%CITATION = doi:10.1103/PhysRevD.58.034016;%%
  %68 citations counted in INSPIRE as of 08 Apr 2019

%\cite{Kniehl:2008zza}
\bibitem{Kniehl:2008zza} 
  B.~A.~Kniehl, G.~Kramer, I.~Schienbein and H.~Spiesberger,
  %``Finite-mass effects on inclusive $B$ meson hadroproduction,''
  Phys.\ Rev.\ D {\bf 77}, 014011 (2008)
  doi:10.1103/PhysRevD.77.014011
  [arXiv:0705.4392 [hep-ph]].
  %%CITATION = doi:10.1103/PhysRevD.77.014011;%%
  %102 citations counted in INSPIRE as of 08 Apr 2019

%\cite{Heister:2001jg}
\bibitem{Heister:2001jg} 
  A.~Heister {\it et al.} [ALEPH Collaboration],
  %``Study of the fragmentation of b quarks into B mesons at the Z peak,''
  Phys.\ Lett.\ B {\bf 512}, 30 (2001)
  doi:10.1016/S0370-2693(01)00690-6
  [hep-ex/0106051].
  %%CITATION = doi:10.1016/S0370-2693(01)00690-6;%%
  %201 citations counted in INSPIRE as of 08 Apr 2019

%\cite{Abbiendi:2002vt}
\bibitem{Abbiendi:2002vt} 
  G.~Abbiendi {\it et al.} [OPAL Collaboration],
  %``Inclusive analysis of the b quark fragmentation function in Z decays at LEP,''
  Eur.\ Phys.\ J.\ C {\bf 29}, 463 (2003)
  doi:10.1140/epjc/s2003-01229-x
  [hep-ex/0210031].
  %%CITATION = doi:10.1140/epjc/s2003-01229-x;%%
  %136 citations counted in INSPIRE as of 08 Apr 2019

%\cite{Abe:2002iq}
\bibitem{Abe:2002iq} 
  K.~Abe {\it et al.} [SLD Collaboration],
  %``Measurement of the b quark fragmentation function in Z0 decays,''
  Phys.\ Rev.\ D {\bf 65}, 092006 (2002)
  Erratum: [Phys.\ Rev.\ D {\bf 66}, 079905 (2002)]
  doi:10.1103/PhysRevD.66.079905, 10.1103/PhysRevD.65.092006
  [hep-ex/0202031].
  %%CITATION = doi:10.1103/PhysRevD.66.079905, 10.1103/PhysRevD.65.092006;%%
  %128 citations counted in INSPIRE as of 08 Apr 2019

%\cite{DELPHI:2011aa}
\bibitem{DELPHI:2011aa} 
  J.~Abdallah {\it et al.} [DELPHI Collaboration],
  %``A study of the b-quark fragmentation function with the DELPHI detector at LEP I and an averaged distribution obtained at the Z Pole,''
  Eur.\ Phys.\ J.\ C {\bf 71}, 1557 (2011)
  doi:10.1140/epjc/s10052-011-1557-x
  [arXiv:1102.4748 [hep-ex]].
  %%CITATION = doi:10.1140/epjc/s10052-011-1557-x;%%
  %54 citations counted in INSPIRE as of 08 Apr 2019

%\cite{Pumplin:2000vx}
\bibitem{Pumplin:2000vx} 
  J.~Pumplin, D.~R.~Stump and W.~K.~Tung,
  %``Multivariate fitting and the error matrix in global analysis of data,''
  Phys.\ Rev.\ D {\bf 65}, 014011 (2001)
  doi:10.1103/PhysRevD.65.014011
  [hep-ph/0008191].
  %%CITATION = doi:10.1103/PhysRevD.65.014011;%%
  %118 citations counted in INSPIRE as of 08 Apr 2019

%\cite{Chetyrkin:1979bj}
\bibitem{Chetyrkin:1979bj} 
  K.~G.~Chetyrkin, A.~L.~Kataev and F.~V.~Tkachov,
  %``Higher Order Corrections to Sigma-t (e+ e- ---> Hadrons) in Quantum Chromodynamics,''
  Phys.\ Lett.\  {\bf 85B}, 277 (1979).
  doi:10.1016/0370-2693(79)90596-3
  %%CITATION = doi:10.1016/0370-2693(79)90596-3;%%
  %620 citations counted in INSPIRE as of 08 Apr 2019

%\cite{Kartvelishvili:1985ac}
\bibitem{Kartvelishvili:1985ac} 
  V.~G.~Kartvelishvili and A.~K.~Likhoded,
  %``Structure Functions and Leptonic Widths of Heavy Mesons,''
  Yad.\ Fiz.\  {\bf 42}, 1306 (1985)
  [Sov.\ J.\ Nucl.\ Phys.\  {\bf 42}, 823 (1985)].
  %%CITATION = YAFIA,42,1306;%%
  %25 citations counted in INSPIRE as of 08 Apr 2019

%\cite{Tanabashi:2018oca}
\bibitem{Tanabashi:2018oca} 
  M.~Tanabashi {\it et al.} [Particle Data Group],
  %``Review of Particle Physics,''
  Phys.\ Rev.\ D {\bf 98}, no. 3, 030001 (2018).
  doi:10.1103/PhysRevD.98.030001
  %%CITATION = doi:10.1103/PhysRevD.98.030001;%%
  %1352 citations counted in INSPIRE as of 08 Apr 2019

%\cite{Stump:2001gu}
\bibitem{Stump:2001gu} 
  D.~Stump, J.~Pumplin, R.~Brock, D.~Casey, J.~Huston, J.~Kalk, H.~L.~Lai and W.~K.~Tung,
  %``Uncertainties of predictions from parton distribution functions. 1. The Lagrange multiplier method,''
  Phys.\ Rev.\ D {\bf 65}, 014012 (2001)
  doi:10.1103/PhysRevD.65.014012
  [hep-ph/0101051].
  %%CITATION = doi:10.1103/PhysRevD.65.014012;%%
  %257 citations counted in INSPIRE as of 08 Apr 2019

%\cite{Blumlein:2006be}
\bibitem{Blumlein:2006be} 
  J.~Blumlein, H.~Bottcher and A.~Guffanti,
  %``Non-singlet QCD analysis of deep inelastic world data at O(alpha(s)**3),''
  Nucl.\ Phys.\ B {\bf 774}, 182 (2007)
  doi:10.1016/j.nuclphysb.2007.03.035
  [hep-ph/0607200].
  %%CITATION = doi:10.1016/j.nuclphysb.2007.03.035;%%
  %151 citations counted in INSPIRE as of 08 Apr 2019

%\cite{James:1975dr}
\bibitem{James:1975dr} 
  F.~James and M.~Roos,
  %``Minuit: A System for Function Minimization and Analysis of the Parameter Errors and Correlations,''
  Comput.\ Phys.\ Commun.\  {\bf 10}, 343 (1975).
  doi:10.1016/0010-4655(75)90039-9
  %%CITATION = doi:10.1016/0010-4655(75)90039-9;%%
  %2145 citations counted in INSPIRE as of 08 Apr 2019

%\cite{Bertone:2013vaa}
\bibitem{Bertone:2013vaa} 
  V.~Bertone, S.~Carrazza and J.~Rojo,
  %``APFEL: A PDF Evolution Library with QED corrections,''
  Comput.\ Phys.\ Commun.\  {\bf 185}, 1647 (2014)
  doi:10.1016/j.cpc.2014.03.007
  [arXiv:1310.1394 [hep-ph]].
  %%CITATION = doi:10.1016/j.cpc.2014.03.007;%%
  %131 citations counted in INSPIRE as of 08 Apr 2019
  
%\cite{Martin:2009iq}
\bibitem{Martin:2009iq} 
  A.~D.~Martin, W.~J.~Stirling, R.~S.~Thorne and G.~Watt,
  %``Parton distributions for the LHC,''
  Eur.\ Phys.\ J.\ C {\bf 63}, 189 (2009)
  doi:10.1140/epjc/s10052-009-1072-5
  [arXiv:0901.0002 [hep-ph]].
  %%CITATION = doi:10.1140/epjc/s10052-009-1072-5;%%
  %4420 citations counted in INSPIRE as of 08 Apr 2019

%\cite{Corcella:2001hz}
\bibitem{Corcella:2001hz} 
  G.~Corcella and A.~D.~Mitov,
  %``Bottom quark fragmentation in top quark decay,''
  Nucl.\ Phys.\ B {\bf 623}, 247 (2002)
  doi:10.1016/S0550-3213(01)00639-3
  [hep-ph/0110319].
  %%CITATION = doi:10.1016/S0550-3213(01)00639-3;%%
  %45 citations counted in INSPIRE as of 08 Apr 2019

%\cite{Cacciari:2002re}
\bibitem{Cacciari:2002re} 
  M.~Cacciari, G.~Corcella and A.~D.~Mitov,
  %``Soft gluon resummation for bottom fragmentation in top quark decay,''
  JHEP {\bf 0212}, 015 (2002)
  doi:10.1088/1126-6708/2002/12/015
  [hep-ph/0209204].
  %%CITATION = doi:10.1088/1126-6708/2002/12/015;%%
  %33 citations counted in INSPIRE as of 08 Apr 2019

 %\cite{Kniehl:2012mn}
\bibitem{Kniehl:2012mn} 
  B.~A.~Kniehl, G.~Kramer and S.~M.~Moosavi Nejad,
  %``Bottom-Flavored Hadrons from Top-Quark Decay at Next-to-Leading order in the General-Mass Variable-Flavor-Number Scheme,''
  Nucl.\ Phys.\ B {\bf 862}, 720 (2012)
  doi:10.1016/j.nuclphysb.2012.05.008
  [arXiv:1205.2528 [hep-ph]].
  %%CITATION = doi:10.1016/j.nuclphysb.2012.05.008;%%
  %18 citations counted in INSPIRE as of 08 Apr 2019

 %\cite{Fischer:1998gsa}
\bibitem{Fischer:1998gsa} 
  M.~Fischer, S.~Groote, J.~G.~Korner, M.~C.~Mauser and B.~Lampe,
  %``Polarized top decay into polarized W: t(polarized) ---> W(polarized) + b at O(alpha-s),''
  Phys.\ Lett.\ B {\bf 451}, 406 (1999)
  doi:10.1016/S0370-2693(99)00194-X
  [hep-ph/9811482].
  %%CITATION = doi:10.1016/S0370-2693(99)00194-X;%%
  %59 citations counted in INSPIRE as of 17 Apr 2019
  
%\cite{Fischer:2001gp}
\bibitem{Fischer:2001gp} 
  M.~Fischer, S.~Groote, J.~G.~Korner and M.~C.~Mauser,
  %``Complete angular analysis of polarized top decay at O(alpha($s$) ),''
  Phys.\ Rev.\ D {\bf 65}, 054036 (2002)
  doi:10.1103/PhysRevD.65.054036
  [hep-ph/0101322].
  %%CITATION = doi:10.1103/PhysRevD.65.054036;%%
  %83 citations counted in INSPIRE as of 17 Apr 2019

 %\cite{Nejad:2013fba}
\bibitem{Nejad:2013fba} 
  S.~M.~Moosavi Nejad,
  %``Energy spectrum of bottom- and charmed-flavored mesons from polarized top quark decay $t(↑)→W^++B/D+X$ at $O(α_s)$,''
  Phys.\ Rev.\ D {\bf 88}, no. 9, 094011 (2013)
  doi:10.1103/PhysRevD.88.094011
  [arXiv:1310.5686 [hep-ph]].
  %%CITATION = doi:10.1103/PhysRevD.88.094011;%%
  %13 citations counted in INSPIRE as of 08 Apr 2019
  
 %\cite{Nejad:2014sla}
\bibitem{Nejad:2014sla}
S.~M.~Moosavi Nejad and M.~Balali,
%``Angular analysis of polarized top quark decay into $B$-mesons in two different helicity systems,''
Phys.\ Rev.\ D {\bf 90} (2014) no.11,  114017
Erratum: [Phys.\ Rev.\ D {\bf 93} (2016) no.11,  119904]
doi:10.1103/PhysRevD.90.114017, 10.1103/PhysRevD.93.119904
[arXiv:1409.1389 [hep-ph]].
%%CITATION = doi:10.1103/PhysRevD.90.114017, 10.1103/PhysRevD.93.119904;%%
%6 citations counted in INSPIRE as of 15 Apr 2019  
  
%\cite{Kramer:2001gd}
\bibitem{Kramer:2001gd} 
  G.~Kramer and H.~Spiesberger,
  %``Inclusive D* production in photon photon collisions at next-to-leading order QCD,''
  Eur.\ Phys.\ J.\ C {\bf 22}, 289 (2001)
  doi:10.1007/s100520100805
  [hep-ph/0109167].
  %%CITATION = doi:10.1007/s100520100805;%%
  %48 citations counted in INSPIRE as of 08 Apr 2019

%\cite{Kniehl:2004fy}
\bibitem{Kniehl:2004fy} 
  B.~A.~Kniehl, G.~Kramer, I.~Schienbein and H.~Spiesberger,
  %``Inclusive D*+- production in p anti-p collisions with massive charm quarks,''
  Phys.\ Rev.\ D {\bf 71}, 014018 (2005)
  doi:10.1103/PhysRevD.71.014018
  [hep-ph/0410289].
  %%CITATION = doi:10.1103/PhysRevD.71.014018;%%
  %122 citations counted in INSPIRE as of 08 Apr 2019

 %\cite{Kniehl:2005mk}
\bibitem{Kniehl:2005mk} 
  B.~A.~Kniehl, G.~Kramer, I.~Schienbein and H.~Spiesberger,
  %``Collinear subtractions in hadroproduction of heavy quarks,''
  Eur.\ Phys.\ J.\ C {\bf 41}, 199 (2005)
  doi:10.1140/epjc/s2005-02200-7
  [hep-ph/0502194].
  %%CITATION = doi:10.1140/epjc/s2005-02200-7;%%
  %103 citations counted in INSPIRE as of 08 Apr 2019

 %\cite{Kniehl:2005ej}
\bibitem{Kniehl:2005ej} 
  B.~A.~Kniehl, G.~Kramer, I.~Schienbein and H.~Spiesberger,
  %``Reconciling open charm production at the Fermilab Tevatron with QCD,''
  Phys.\ Rev.\ Lett.\  {\bf 96}, 012001 (2006)
  doi:10.1103/PhysRevLett.96.012001
  [hep-ph/0508129].
  %%CITATION = doi:10.1103/PhysRevLett.96.012001;%%
  %105 citations counted in INSPIRE as of 08 Apr 2019

%\cite{Collins:1998rz}
\bibitem{Collins:1998rz} 
  J.~C.~Collins,
  %``Hard scattering factorization with heavy quarks: A General treatment,''
  Phys.\ Rev.\ D {\bf 58}, 094002 (1998)
  doi:10.1103/PhysRevD.58.094002
  [hep-ph/9806259].
  %%CITATION = doi:10.1103/PhysRevD.58.094002;%%
  %278 citations counted in INSPIRE as of 08 Apr 2019

%\cite{Kneesch:2007ey}
\bibitem{Kneesch:2007ey} 
  T.~Kneesch, B.~A.~Kniehl, G.~Kramer and I.~Schienbein,
  %``Charmed-meson fragmentation functions with finite-mass corrections,''
  Nucl.\ Phys.\ B {\bf 799}, 34 (2008)
  doi:10.1016/j.nuclphysb.2008.02.015
  [arXiv:0712.0481 [hep-ph]].
  %%CITATION = doi:10.1016/j.nuclphysb.2008.02.015;%%
  %124 citations counted in INSPIRE as of 08 Apr 2019

%\cite{Nejad:2016epx}
\bibitem{Nejad:2016epx} 
  S.~M.~Moosavi Nejad and M.~Balali,
  %``Hadron energy spectrum in polarized top quark decays considering the effects of hadron and bottom quark masses,''
  Eur.\ Phys.\ J.\ C {\bf 76}, no. 3, 173 (2016)
  doi:10.1140/epjc/s10052-016-4017-9
  [arXiv:1602.05322 [hep-ph]].
  %%CITATION = doi:10.1140/epjc/s10052-016-4017-9;%%
  %4 citations counted in INSPIRE as of 08 Apr 2019

%\cite{Albino:2005gd}
\bibitem{Albino:2005gd} 
  S.~Albino, B.~A.~Kniehl, G.~Kramer and W.~Ochs,
  %``Resummation of soft gluon logarithms in the DGLAP evolution of fragmentation functions,''
  Phys.\ Rev.\ D {\bf 73}, 054020 (2006)
  doi:10.1103/PhysRevD.73.054020
  [hep-ph/0510319].
  %%CITATION = doi:10.1103/PhysRevD.73.054020;%%
  %33 citations counted in INSPIRE as of 08 Apr 2019

%\cite{Albino:2006wz}
\bibitem{Albino:2006wz} 
  S.~Albino, B.~A.~Kniehl, G.~Kramer and C.~Sandoval,
  %``Confronting fragmentation function universality with single hadron inclusive production at HERA and e+ e- colliders,''
  Phys.\ Rev.\ D {\bf 75}, 034018 (2007)
  doi:10.1103/PhysRevD.75.034018
  [hep-ph/0611029].
  %%CITATION = doi:10.1103/PhysRevD.75.034018;%%
  %25 citations counted in INSPIRE as of 08 Apr 2019

%\cite{Albino:2008fy}
\bibitem{Albino:2008fy} 
  S.~Albino, B.~A.~Kniehl and G.~Kramer,
  %``AKK Update: Improvements from New Theoretical Input and Experimental Data,''
  Nucl.\ Phys.\ B {\bf 803}, 42 (2008)
  doi:10.1016/j.nuclphysb.2008.05.017
  [arXiv:0803.2768 [hep-ph]].
  %%CITATION = doi:10.1016/j.nuclphysb.2008.05.017;%%
  %241 citations counted in INSPIRE as of 08 Apr 2019

\end{thebibliography}
\end{document}